\documentclass[useAMS,usenatbib]{mn2e}

\usepackage{graphicx}
\usepackage{threeparttable}

\newcommand{\rmxaa}{Revista Mexicana de Astronomia y Astrofisica}

\newcommand{\pasp}{{\it PASP~\/}}

\newcommand{\apj}{ApJ}
\newcommand{\apjl}{ApJ}
\newcommand{\apjs}{ApJS}
\newcommand{\aap}{A \& A}
\newcommand{\araa}{ARA\&A}

\newcommand{\mnras}{MNRAS}

\newcommand{\nat}{Nature}

\makeatletter
\newcommand*{\rom}[1]{\expandafter\@slowromancap\romannumeral #1@}
\makeatother

\title[Evolution of gas around $\beta$ Pictoris]{A self-consistent model for the evolution of the gas produced in the debris disc of $\beta$ Pictoris}
\author[Q. Kral, M. Wyatt, R. F. Carswell, J. E. Pringle, L. Matr\`a, A. Juh\'asz]{Q. Kral\thanks{E-mail: qkral@ast.cam.ac.uk}, M. Wyatt, R. F. Carswell, J. E.  Pringle, L. Matr\`a, A. Juh\'asz \\
Institute of Astronomy, University of Cambridge, Madingley Road, Cambridge, UK, CB3 0HA}
\begin{document}

\date{Accepted 1928 December 15. Received 1928 December 14; in original form 1928 October 11}

\pagerange{\pageref{firstpage}--\pageref{lastpage}} \pubyear{2002}

\maketitle

\label{firstpage}

\begin{abstract}
This paper presents a self-consistent model for the evolution of gas produced in the debris disc of $\beta$ Pictoris. Our model proposes that atomic carbon and oxygen are created from the photodissociation of CO, which is itself released from volatile-rich bodies in the debris disc due to grain-grain collisions or photodesorption. 
While the CO lasts less than one orbit, the atomic gas evolves by viscous spreading resulting in an accretion disc inside the parent belt and a decretion disc outside. The temperature, ionisation fraction and population levels of carbon and oxygen are followed with the photodissociation region
model Cloudy, which is coupled to a dynamical viscous $\alpha$ model. We present new gas observations of $\beta$ Pic, of C I observed with APEX and O I observed with Herschel, and show that these along with published C II and CO observations can all be explained with this new model.
Our model requires a viscosity $\alpha > 0.1$, similar to that found in sufficiently ionised discs of other astronomical objects; we propose that the magnetorotational instability is at play in this highly ionised and dilute medium. 
This new model can be tested from its predictions for high resolution ALMA observations of C I. We also constrain the water content of the planetesimals in $\beta$ Pic. The scenario proposed here might be at play in all debris discs and this model could 
be used more generally on all discs with C, O or CO detections.

\end{abstract}


\section{Introduction}

The first observation of circumstellar gas around a star with a debris disc dates back to 1975, even before the discovery of debris discs themselves. A spectrum of $\beta$ Pictoris \citep[located at 19.44 $\pm$ 0.05 pc,][]{2007A&A...474..653V} revealed the presence of Ca II in absorption, which was presumably 
of circumstellar origin \citep{1975ApJ...197..137S}. The discovery of circumstellar dust around $\beta$ Pic and a handful of stars by IRAS \citep{1985PASP...97..885A} provided a strong indication of the origin of the gas, and 
ever since gas has been looked for around stars with debris discs. Since the $\beta$ Pictoris disc is observed edge-on, it is a good target in which to detect UV absorption lines in the star's spectrum that are due to circumstellar material 
along the line-of-sight. Elements such as C, O, Na, Mg, Al, Si, S, Ca, Cr, Mn, Fe, Ni have been detected through their UV lines in $\beta$ Pictoris \citep{2006Natur.441..724R}, while observations of hydrogen have provided upper limits
on the column density of H I \citep{1995A&A...301..231F} and H$_2$ \citep{2001Natur.412..706L}. 

One puzzling observation is
that metals that should be strongly affected by radiation pressure (such as Na I) seem to be on Keplerian orbits \citep{2001ApJ...563L..77O}. It was proposed that an overabundance of carbon could explain this behaviour as it is a strong braking agent
\citep{2006ApJ...643..509F}. This is because the metals spend the majority of their time ionised, and while ionised, they couple to the ionised carbon which is in Keplerian rotation as it does not feel radiation pressure.

This prediction was confirmed by a Herschel/HIFI C II emission spectrum, which was used to infer an overabundance of carbon and oxygen by a factor that is
up to 400 times the solar abundance \citep{2014A&A...563A..66C}. However, the origin of this overabundance was unclear. This suggests that either C and O are produced preferentially, or other elements are preferentially depleted \citep{2013ApJ...762..114X}. 

A clue to the origin of the overabundance of C and O in the disc comes from spatially resolved ALMA observations of CO gas orbiting the star. The emission is highly asymmetric, with 30\% of the emission located in a single clump on one side of the star, with kinematics giving the exact 
position of the gas clump at 85 AU from the star \citep{2014Sci...343.1490D}, possibly colocated with a dust clump detected in the mid-IR \citep{2005Natur.433..133T}. The radial distribution of CO is similar to the location of the planetesimal belt inferred from the continuum emission 
to extend 50-150 AU, which is thus thought to be the source of gas. 

We have now reached the point where enough observations are available for $\beta$ Pic to start tackling some important conundrums. The main questions concern the origin of the gas and its subsequent evolution. The molecular gas observed around $\beta$ Pic cannot be primordial as the photodissociation
 timescale of the observed CO is $\sim 120$ years \citep{2014Sci...343.1490D} whereas the system is a lot older \citep[21$\pm$4 Myr,][]{2014MNRAS.438L..11B}. Since the molecular gas is colocated with the dust, it is thought to be secondary, i.e it is released from volatile-rich
solid bodies. CO can be kept on solid bodies at 85AU (even though the sublimation temperature of CO is 20K) as it can be trapped in water ice up to 140K \citep{2003ApJ...583.1058C}. When CO is ejected from solid bodies, it is not expected to recondense on grains as the dust density 
is too low in this disc. Photodesorption \citep{2007A&A...475..755G}, vaporisation of dust grains during high-velocity collisions \citep{2007ApJ...660.1541C}, collisions between volatile-rich comets \citep{2012ApJ...758...77Z}, and sublimating comets \citep{1990A&A...236..202B} are the main 
processes proposed to release gas from solid bodies. 

While the photodissociation of CO must release C and O into the gas disc, it has yet to be determined what this implies for the atomic species, and whether for example it is possible to explain all observations within one self-consistent model. It is expected that the atomic gas would evolve viscously,
however \citet{2014A&A...563A..66C} concluded that the C II spectrum, which is sufficiently resolved to give some information on the radial distribution of ionised carbon, is inconsistent with an accretion disc profile. However, their model included several simplifying assumptions about the thermodynamic state
of the gas disc and about the profile expected for an accretion disc.

In this paper, we develop a thermodynamical model of gas in debris discs and apply it to explain $\beta$ Pictoris gas observations. In this new model, we assume that the atomic elements C and O are produced through 
photodissociation of CO. We then follow the temporal evolution of carbon and oxygen atomic elements assuming that they diffuse viscously. The viscosity is parameterised with an $\alpha$ coefficient as is typical when studying 
gas evolution in protoplanetary discs \citep{1973A&A....24..337S}. The temperature and ionisation state of C and O are worked out using Cloudy \citep{2013RMxAA..49..137F}, which is a PDR-like model. The impinging radiation 
on the gas disc is composed of the stellar and the interstellar radiation fields.

The second section details the numerical model used to model gas observations of $\beta$ Pictoris. The third section presents our results and shows how this model is able to fit the recent C II and O I Herschel observations as well as a new C I non-detection by APEX in a self-consistent way. 
Lastly, we discuss our results in section four.

\section{Numerical Model for gas evolution}\label{model}

To understand the distribution of gas in $\beta$ Pictoris, we develop a numerical model of viscous diffusion of gas in the low density regime expected in debris discs. We assume that CO is being produced in the main belt and quickly photodissociates, 
leading to an input of gaseous C and O at that radius. We model the subsequent evolution of this gaseous component using standard accretion disc physics. The modelled gas sits in a dust disc, which extends from $\sim$ 50 to $\sim$ 150AU \citep{2001A&A...370..447A,2014Sci...343.1490D} and might affect the thermal state of the gas.

Throughout this paper, we choose to use our general model on $\beta$ Pic so that we take the gas injection location to be $R_0=85$ AU, where the bulk of CO is located. The CO mass input rate is estimated to be $1.4 \times 10^{18}$ kg/yr \citep{2014Sci...343.1490D}. For the carbon input rate, 
this corresponds to $\dot{M}=0.1 \mathrm{M}_\oplus$/Myr. We also input the right amount of oxygen 
taking C/O = 1 in number density. We take the radiation field $F$ to be composed of an A6V star $F_\star$ plus the interstellar radiation field $F_i$, which is assumed to be that derived by \citet{1983A&A...128..212M} $F_0$ but multiplied by a constant $X$, to change the amount of UV radiation impinging the gas disc so that $F=F_\star+F_i$, where $F_i=X F_0$. The other free parameter is the viscosity parameter $\alpha$ (defined in subsection \ref{fiduc}).

\subsection{The accretion disc}\label{fiduc}

The gas evolution model developed here is very general and can be applied to a range of situations where gas is injected in a system at a specific location. The model treats the radial evolution of the gas using the standard evolution 
equation for the surface density, but at each time-step solves for the local vertical structure as a one-zone model. This is a gross simplification, but is adequate for an initial investigation.

The evolution of the gas under viscous diffusion is followed using the  equation \citep{1974MNRAS.168..603L,1981ARA&A..19..137P}

\begin{equation}
\label{eqdif}
 \frac{\partial \Sigma}{\partial t} = \frac{3}{R} \frac{\partial}{\partial R} \left( \sqrt{R} \frac{\partial}{\partial R} (\nu \Sigma \sqrt{R})  \right) + \dot{\Sigma}_0 \, \delta(\frac{R}{R_0}-1),
\end{equation}
where $\nu$ is the kinematic viscosity, $\Sigma(R,t)$ the surface density, $R$ the radial variable and $\dot{\Sigma}_0$ the surface density input rate at radius $R_0$. Here $\delta$ is the usual Dirac delta-function. In practice gas input takes 
place over a range of radii $\Delta R_0$ and thus, $\dot{\Sigma}_0$ is related to the mass input rate $\dot{M}$ in the following manner: $\dot{M}=2 \pi R_0 \Delta R_0 \dot{\Sigma}_0$. The value of $\Delta R_0$ is taken to be one radial size
 bin and is not important for our model as soon as $\Delta R_0/R_0 \ll 1$ as the value of $\dot{\Sigma}_0$ is worked out over the same bin size. We treat $\dot{M}$ as a free parameter.

In the absence of any other information we use the standard \citet{1973A&A....24..337S}  $\alpha$-parametrisation for the viscosity. Thus we write
\begin{equation}
\label{nualpha}
 \nu_\alpha = \alpha c_s H,
\end{equation}
where $c_s$ is the sound speed and $H$ the local one-zone disc scale height. 

The value of $\alpha$ and its dependence on disc properties is one of the main uncertainties of current accretion disc theory. Observational evidence from X-ray outbursts and dwarf novae, where discs are fully, or sufficiently,  ionised, suggests high $\alpha$ values ranging from 0.1 to 0.4 \citep{2007MNRAS.376.1740K}.
However, much lower values are expected to prevail in regions of low ionisation \citep{1998ApJ...492L..75G}. Depending on the mechanism producing this viscosity in cool protoplanetary discs, $\langle \alpha \rangle$ produced in numerical simulations can vary between $10^{-4}-1$ \citep{1998RvMP...70....1B,2013ApJ...767...30B}, or can even be essentially zero in the so-called dead zones \citep{2012MNRAS.424.1977L}.

In order to compute the vertical disc scale-height we make the approximation that there is a single temperature $T(R,t)$ which adequately describes the local disc structure. 
We then obtain the disc scale-height $H(R,t)$  where $H=c_s/\Omega$, $c_s$ being the sound speed and $\Omega$ the orbital frequency. The sound speed $c_s$ is fixed by the gas temperature $T$ as 
$c_s=\sqrt{R_g T / \mu}$, with $R_g$ being the ideal gas constant and $\mu$ the mean molecular mass, which can be substantially different from that of protoplanetary discs as we expect most of the mass to be in carbon and oxygen rather than hydrogen.
The gas density $\rho_\mathrm{g}(R,z)$ is assumed to be constant over a distance $H$ in the vertical direction.  Hence, $\Sigma=2 \rho H$.
To convert between $\Sigma$ and the particle number density $n$, we use $n = \rho/(\mu m_p)$, where $m_p$ is the proton mass

\begin{equation}
\label{eqn}
 n = \frac{ \Omega \Sigma}{2 \mu m_p c_s}.
\end{equation}

In our numerical simulations, we need to set the boundary conditions.  We choose an inner radius of $R_\mathrm{min} = 5$AU sufficiently close to the host star that we can follow the disc radii of interest to us. We choose an outer disc radius $R_\mathrm{max} = 1000$AU to be large enough that it has no significant effect on the results.

We assume that all material reaching the inner boundary is accreted onto the star or onto the planet $\beta$ Pic b, and thus set  $\Sigma(R_\mathrm{min})=0$. For convenience, at  the outer boundary, we assume that the material is lost at that point and set  $\Sigma(R_\mathrm{max})=0$.

If we know the radial temperature profile of the disc, we are now in a position to compute its evolution.
In the next subsection we compute the evolution for a given power law temperature profile. Later, in contrast,  we compute the local temperature structure in a more physical manner.



\subsection{Evolution to the steady state of an $\alpha$ disc with fixed temperature profile}\label{ssalpha}


To understand the evolution of an $\alpha$ disc, we first assume that the temperature profile is a power law defined by $T=T_0 (R/R_0)^{-\gamma}$, where $T_0$ = 60K and $\gamma=0.5$. 

\begin{figure}
 \includegraphics[width=8.5cm]{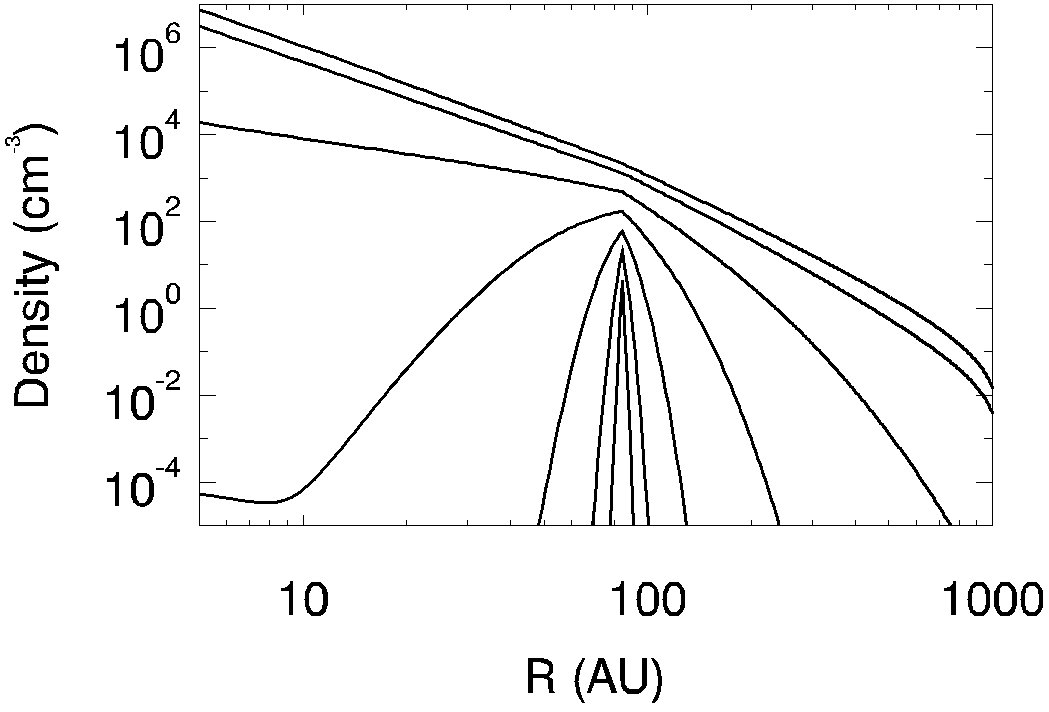}
   \caption{\label{figA} Density profile at different epochs logarithmically spaced between 2 years and $5 \times 10^5$ years (each curve is separated by a factor $\sim$ 8 in time). The temperature is assumed to have a power law dependence imposed (see subsection \ref{ssalpha}).  The last time-step at 500 000 years shows the expected steady state accretion and decretion profiles.}
\end{figure}

Fig.~\ref{figA} shows the evolution resulting from gas injected at $R_0=85$AU at a constant rate $\dot{M} = 0.1 \mathrm{M}_\oplus$/Myr. We take $\alpha=0.5$. For this computation we have taken $R_{\rm min} = 5$AU and $R_{\rm max} = 1056$AU. We use a radial grid of 400 points equally spaced in $\sqrt{R}$ \citep{1981MNRAS.194..967B}. At time $t=0$, mass is injected steadily and the number density starts growing, 
creating a spike around 85AU.
At the same time, it is viscously evolving and creates an accretion disc inwards and a decretion disc outwards. By $t \sim 5 \times 10^5$ years, steady state is reached 
for the accretion and decretion discs. We note that the fraction $f$ of matter injected at radius $R_0$ that is lost at the outer radius $R_{\rm max}$ is given, for $R_{\rm min} \ll R_0$, by $f = (R_0/R_{\rm max})^{1/2} \sim 0.3$. Thus, for the parameters we have chosen, 70 percent of the input material is accreted onto the central star.

The timescale to reach steady state for the disc should be close to the viscous timescale at $R_0$, i.e $t_\nu \sim R_0^2/\nu(R_0)$, which is $\sim 10^5$ years for the parameters of the simulation in Fig.~\ref{figA}.

\subsection{The full model}

To obtain a more realistic model of the radial temperature behaviour we need to consider the thermal equilibrium of the gas at each radius.
 
\begin{figure}
 \includegraphics[width=8.5cm]{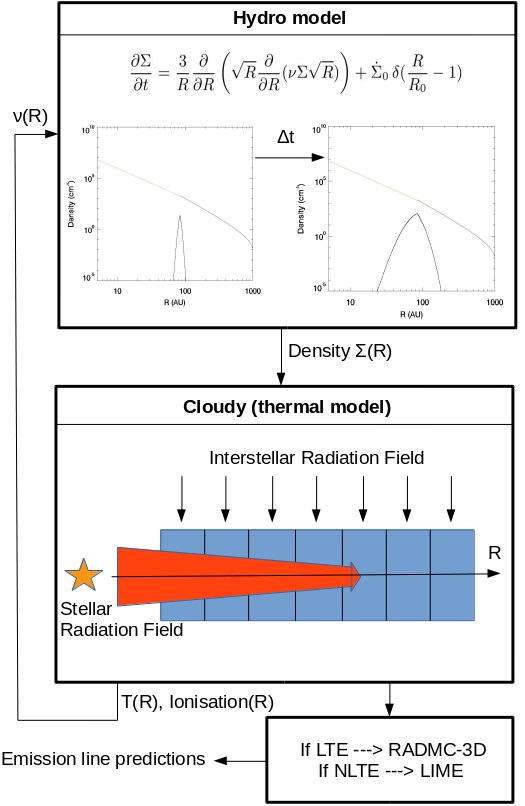}
   \caption{\label{figdiagram} Diagram explaining the coupling between the dynamical and thermal models. The upper box shows the hydro model that evolves the density $\Sigma(R)$ through solving the diffusion equation in time over $\Delta t$. The new density 
is then passed to Cloudy, the thermal model, which solves for the new temperature $T(R)$ and ionisation fractions in each cell shown on the diagram in blue. It takes account of the photons coming from the central star (and their extinction along $R$)
as well as the interstellar radiation field coming from the top. Hence, emission lines can be predicted using either RADMC-3D in LTE or LIME in NLTE. The viscosity $\nu(R)$ can be worked out from the temperature profile and is input back into the hydro model to start the next timestep.}
\end{figure}

We model the thermal state of the gas using Cloudy \citep{2013RMxAA..49..137F}, a spectral synthesis code used to study gas clouds under different conditions.  We are then able to follow the viscous evolution of the gas within the disc with a time-dependent, and locally determined, temperature profile, which depends on $R$, with a more complex dependence than a simple power law. Hence $\nu$ also depends on $R$ and is reinjected into Eq.~\ref{eqdif} to compute the diffusion of the gas for the next time step. To compute the lines of interest for a given timestep, we use RADMC-3D when LTE can be assumed or LIME otherwise (see subsection \ref{emlinec} for more details). 
Fig.~\ref{figdiagram} explains schematically how the model works.
   
We recall that the model has four free parameters, which are the input radial location $R_0$, the mass input rate $\dot{M}$, the viscosity parameter $\alpha$, taken to be constant and the radiation field impinging the disc $F$, 
whose main components are the radiation from the central star $F_\star$ and the external interstellar radiation field $F_i$. Comparing these, we shall find that the $F_\star$ dominates close to the centre of the disc at $R < 30$AU if the medium is optically thin in the continuum (and if a standard
interstellar radiation field is considered $F_i=F_0$), 
and $F_i$ outside that radius where most of the emission lines of interest to us are produced.

\subsubsection{A more realistic temperature profile}

The temperature profile depends on the local heating and cooling mechanisms, which in turn depend on the atomic type, densities, radiation and density of colliders. To work out the temperature self-consistently and
take into account all the physics at play we use Cloudy, a PDR-like spectral synthesis code that works out the gas state depending on its density, composition and incoming radiation. 

Cloudy is a 1D model but as the interstellar radiation field is coming from every direction, it would not be correct to model our gas disc as an horizontal slab. Indeed, if the interstellar radiation field was only imposed from the inner parts of the system,
where the star is located, some radiation would be blocked due to continuum optical thickness of the gas disc in the inner parts, and would not reach the outer parts, which is not physical. To have a good representation
of the physical problem, we sliced our disc into $N_b = 35$ annuli which are spaced using the equally spaced radial variable $X = \sqrt{R}$ along the radial direction in the range 5 AU $ < R < $ 1000 AU (thus we have $\Delta X = 0.84$ and $\Delta R/R < 0.75$) and impose the flux from the top (see Fig.~\ref{figdiagram}). Then, the thermal state of each slab is worked out at a specific $R$ by the Cloudy model along the $z$ direction. We
assume a constant density profile along $z$ and iterate a few times with Cloudy to adjust the scale height to the actual temperature. The stellar flux is added to each annulus by combining it with the incoming interstellar radiation field (i.e in the model it appears to arrive from the top). That stellar flux has been corrected for attenuation in interior annuli by summing up optical depths; the stellar flux also has a $1/R^2$ dependence due to the geometry. The optical thickness to FUV radiation in the vertical direction could be a problem as it would unphysically block stellar radiation. However, the optical depth in the vertical direction is always smaller than in the horizontal direction as the gas density extends over $H$ in the $z$ direction
and over $R>H$ in the radial direction and the density increases inwards. This ensures that if the continuum optical thickness is greater than 1 in the vertical direction, it will always be greater in the radial direction and the stellar UV flux would never have
made it to the midplane location anyway. The different slabs are then gathered and the solution is reinjected into the dynamical model (Eq.~\ref{eqdif}) to work out the viscous evolution for the next timestep.

\subsubsection{Composition of the gas}\label{compnum}

The gas in debris disc systems does not seem to be primordial and to some extents reflects the composition of the solid bodies composing the debris disc from which the gas was released. 

In addition to carbon and oxygen, metals have also been observed around $\beta$ Pic \citep[e.g.,][]{2012A&A...544A.134N}. We include these metals in our Cloudy thermal model using a solar abundance. These metals are found not to affect the thermal state of the gas in our $\beta$ Pic model but are included, in case they have any thermal effects in a different regime of the parameter space.

Hydrogen and helium are found to be significantly depleted compared to solar metal abundance \citep{2010ApJ...720..923Z}. 

As we assume that carbon and oxygen come from photodissociation of CO, we fix $\mathrm{O}/\mathrm{C} = 1$. In $\beta$ Pic, carbon and oxygen are overabundant by a factor $\sim$ 400 with respect to other species \citep{2013ApJ...762..114X,2014A&A...563A..66C}. 
Thus, for the metals $X$ (other than C and O), we assume that [$\mathrm{C}/\mathrm{X}$]=$\log_{10}(N_C/N_X)-\log_{10}(N_C/N_X)_{\rm solar}$=2.6, which corresponds to $N_C/N_X=400\,(N_C/N_X)_{\rm solar}$. The gas disc is assumed to be depleted in 
hydrogen such that [$\mathrm{H}/\mathrm{X}$]=-3 and Helium is absent. We note that we tested that adding hydrogen in the system up to $\mathrm{H}/\mathrm{C}$=3 in number density, when C is 400 times solar, had no impact on heating/cooling processes.

Since carbon and oxygen are by far the dominant species in the disc, the mean molecular weight is $\mu = 14$.

Later, we assume that some more oxygen or hydrogen could arise from water photodissociation. There are no observations that fix the amount of water in this system at the moment. However, we find in section \ref{Hline} that increasing the amount of H and O in this way provides a better fit to the observations.

\subsubsection{The dust disc}\label{dustdisc}

As $\beta$ Pic hosts a debris disc, dust is added in our model to analyse the thermal effects it might have. We determine the effect of photoelectric heating \citep[assuming a standard $q$=-3.5 size distribution, e.g.,][]{2013A&A...558A.121K} on the temperature within the disc using Cloudy. 
The dust optical depth profile is taken to be the same as in \citet{2010ApJ...720..923Z}

\begin{equation}
\label{taudust}
 \tau_d(R)=\frac{\sqrt{2} \, \tau_0}{\sqrt{\left(\frac{R}{R_1}\right)^{-\gamma_1}+\left(\frac{R}{R_1}\right)^{\gamma_2}}},
\end{equation}

\noindent where $\gamma_1=4$, $\gamma_2=6$, and $\tau_0=2 \times 10^{-3}$ is the optical depth value at $R_1=120$ AU. These values are empirically determined to fit $\beta$ Pic dust scattered light observations with HST/STIS \citep{2000ApJ...539..435H}. We also use this profile to compute the dust radiation field when using NLTE calculations of population levels.

\subsubsection{Radiation field}\label{radnum}

The main mechanism that heats the gas is the incoming UV flux. The UV radiation field is fairly strong and can affect the electronic states within atoms, ionise them 
or even photodissociate molecules. The radiation field in our model consists of:

\begin{itemize}
 \item $F_\star$, the stellar flux where a \citet{2004astro.ph..5087C} ATLAS stellar atmosphere model is used,
 \item $F_i$, the interstellar radiation field (IRF) $F_0$ with a \citet{1983A&A...128..212M} prescription times a constant $X$.
 \item The dust radiation field using the prescription given in subsection \ref{dustdisc} above (used for NLTE calculations).
\end{itemize}

\noindent Other possible components, such as cosmic ray heating and the cosmic microwave background, are inconsequential.

The stellar flux and IRF both provide UV photons that are energetic enough to ionise some atomic species in the gas; in particular neutral carbon that has an ionisation potential (IP) equal to 11.26eV. 
The ionisation fraction of the modelled gas is thus very sensitive to the incoming UV flux at energies higher than 10eV. Also, due to the Lyman break, there is a strong cut-off in both the stellar and interstellar radiation flux when the energy is greater
than 13.6eV (912 \AA). 

\subsubsection{Emission lines}\label{emlinec}

To calculate observables (e.g. images, spectra) we combined our Cloudy models with the line radiative transfer codes RADMC-3D \citep{2012ascl.soft02015D} and LIME \citep{2010A&A...523A..25B}. 
Both codes take account of the optical thickness of lines in each direction and for both codes we use the LAMDA database \citep{2005A&A...432..369S} to set the Einstein coefficients, transition energies or collision rates. 
For C I and O I, the optical depth can be a lot higher than unity whilst for C II it only reaches values close to unity in the innermost parts. 

In RADMC-3D we interpolate Cloudy outputs, 
 and use spherical coordinates in 3D on a grid logarithmic in $r$ with ($r$,$\theta$,$\phi$)=(300x50x50). 
We assume that the axisymmetric keplerian gas disc axis is inclined at 88 degrees to the line-of-sight to the observer and has a position angle of 29 degrees \citep{2014Sci...343.1490D}. LTE is a good assumption for C I and C II as their critical densities are equal to $3.9 (T/100K)^{-0.13}$ cm$^{-3}$ and $8.7 (T/100K)^{0.5}$ cm$^{-3}$, whilst the electron density (main collider) derived by our model (see section \ref{predi}) or by previous studies \citep[e.g.][]{2014A&A...563A..66C} are greater than 100 cm$^{-3}$. 

For the O I
line, the critical density is higher and equal to $6.3 \times 10^3 (T/100K)^{-0.03}$ cm$^{-3}$ \citep{1989ApJ...342..306H} so that an NLTE approach is necessary because our model predicts an electron number density that is always much below than this critical density (see Fig.~\ref{figdens}). Instead of using approximations in RADMC-3D (such as the large velocity gradient method) we decided to make the full calculations
using LIME; RADMC-3D also does not yet include radiative exchange with the dust continuum radiation field, which proves to be important for the O I line. We used 61 channels with a resolution of 0.63 km/s and 600 pixels (along the x and y-axes) to produce the data cube with a spatial resolution of 0.05'' at the distance of $\beta$ Pic. 
The LIME simulations included emission from the dust (see subsection \ref{dustdisc}) as it emits in the far-IR at a level $\sim$ 17Jy.

We find that the continuum optical depth in the FUV is dominated by C I ionisation. The ionisation cross-section is fairly simple for carbon as it does not vary with frequency over the restricted range [11.26, 13.6] eV and is equal to $\sigma_{\rm ionC}=1.6 \times 10^{-17}$ cm$^2$ \citep{1988ASSL..146...49V} from 11.26eV (ionisation potential
of carbon) to 13.6eV (Lyman break).



\subsubsection{Main thermal mechanisms}\label{mechanism}

The main ionising radiation in our simulations is the UV flux from the IRF, which yields a high ionisation fraction for the carbon. 
The rate of heating compared to cooling sets the gas temperature, which determines both the dynamics of the gas as well as the emission line intensities.
The main heating process
in the Cloudy simulations is photoionisation (PI) of atoms (mainly carbon), with a small contribution from the photoelectric heating (PEH) on dust grains.
The main heating mechanism is different from that assumed in previous studies \citep[e.g.][]{2001A&A...373..641K,2007ApJ...655..528B,2010ApJ...720..923Z} because the overabundance of carbon was not known before the Herschel observations (see subsection \ref{dustbp} for more details).
The main coolant in our model is through the C II fine structure line at 157.7$\mu$m \citep[which is also found by][]{2010ApJ...720..923Z}. 
Other coolants are
negligible. For instance, cooling by CO and CH 
vibrational/rotational transitions is negligible because these molecules are quickly photodissociated; cooling via Ly $\alpha$ emission is also possible, but in our model the gas disc is partially depleted in hydrogen and moreover this 
effect becomes important only at high temperatures (T $>$ 5000K).

\section{Modelling $\beta$ Pic gas observations}\label{modellingbetapic}


The model described in section \ref{model} is applied to the $\beta$ Pic observations described in subsection \ref{obsc} to see whether we can fit these in a self-consistent manner. 



\subsection{Observations used to fit our model}\label{obsc}

In this subsection we detail the four main observations that are used within the paper to fit our model that are summarised in Table~\ref{tab1}. Two of them are new observations presented here for the first time.

\begin{table}
    \caption{Integrated emission flux for the observations used in this paper.}
\begin{center}
\begin{tabular}{|l|c|c|}
  \hline \hline
  Element & Central & Flux \\
   & Wavelength ($\mu$m) & (Jy km/s) \\
  \hline
  CO & 867.5 & 6.6$\pm$0.7 \\
  C II & 157.7 & 372$\pm$10 \\
  C I & 609.7 & $<$ 14 \\
  O I & 63.18 & 110$\pm$15 \\
  \hline
\label{tab1}
\end{tabular}
\end{center}
\end{table}

\begin{itemize}

\item We use the ALMA J=3-2 $^{12}$CO \citep[observed at 867.5$\mu$m,][]{2014Sci...343.1490D} to fix two of our free parameters. The resolution was $\sim$ 12AU (using 27 antennas with projected baseline lengths from 15 to 380m) and the spectral resolution 0.85km/s.
The resulting CO image shows a broad belt of CO gas from 50 to 150AU (colocated with the parent belt of solid bodies) and a peak around 85AU. No gas emission is observed inside 50AU. From the estimates of the total mass of 
CO and assuming that CO is destroyed by photodissociation in 120 years, they derive the CO input rate in the system $\sim 1.4\times 10^{18}$ kg/yr. Accordingly, we define $\dot{M}_0 = 0.1$ M$_\oplus$/Myr and $R_0=$85 AU but allow $\dot{M}$ to vary from $\dot{M}_0$ by a factor 10.

\item \textit{Herschel} HIFI \citep{2010A&A...518L...6D} observations of the C II $^2$P$_{3/2}$-$^2$P$_{1/2}$ transition at 157.7 $\mu$m 
were discussed in detail in \citet{2014A&A...563A..66C}, and the spectrum (presented in red in Fig.~\ref{figCII}) was kindly provided to us by the authors. We use the horizontal polarisation beam and binned the channels to a width of 0.63km/s. The total emission line flux is estimated to be $2.36 \pm 0.06 \times 10^{-14}$ erg/s/cm$^2$ ($\sim$ 372 Jy km/s) with an additional calibration uncertainty of $\sim$ 10\% \citep{2012A&A...537A..17R}.

\begin{figure}
\centering
   \includegraphics[width=8.5cm]{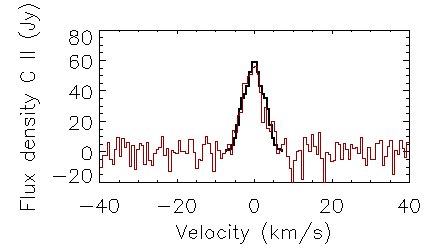}
   \caption{\label{figCII} C II emission line profile predicted for our best-fit model (black) compared to the Herschel observation (red).}
\end{figure}

\item The CI $^3$P$_1$-$^3$P$_0$ transition at 492.16 GHz was observed on 31st August 2015 
with the APEX-3 Swedish Heterodyne Facility Instrument (SHeFI) receiver mounted on the Atacama Pathfinder EXperiment (APEX) telescope, as part of observing program 096.F-9328. Absolute calibration was performed using the chopper wheel method \citep{1976ApJS...30..247U}, 
and standard pointing measurements were carried out to ensure a pointing accuracy of $<$1.5" RMS. The telescope beam at 492 GHz is 12.7" in size (FWHM), and the spectral resolution obtained with the XFFTS spectrometer was 0.046 km/s. 
We spent a total of 20.1 minutes on source in good observing conditions (PWV $\sim$0.5 mm), allowing us to reach an RMS noise level of $\sim$139 mK for a 0.046 km/s channel. The post-processing simply consisted of visual inspection and flagging 
of individual spectra, which were then averaged together with a weight equal to the reciprocal of the system temperature in each channel. We then applied a simple polynomial baseline subtraction around the spectral region of interest to remove any remaining 
background signal, and applied spectral smoothing to increase the signal-to-noise ratio (SNR) on the CI line. The resulting spectrum is shown in red in Fig.~\ref{figCI}. The APEX non-detection imposes that the C I integrated flux at 609 $\mu$m must be smaller than 14 Jy km/s.

\begin{figure}
\centering
   \includegraphics[width=7.5cm]{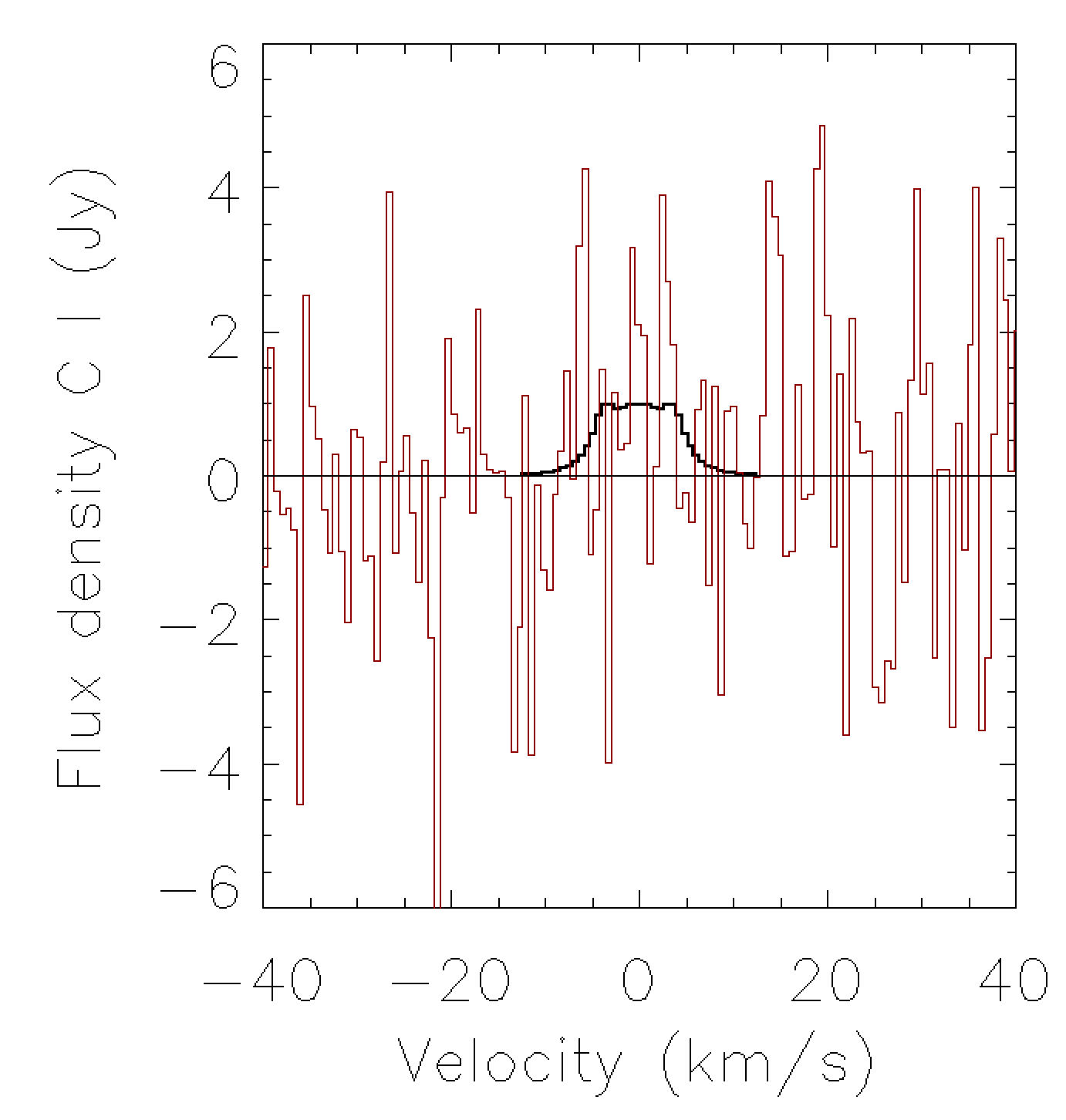}
   \caption{\label{figCI} APEX C I emission line profile observed (non-detection in red) and predicted for our best-fit model (in black).}
\end{figure}


\item We also retrieved archival \textit{Herschel} \citep{2010A&A...518L...1P} observations of the OI $^3$P$_1$-$^3$P$_2$ transition at 63.18 $\mu$m from the \textit{Herschel} Science Archive. 
These were carried out on 22nd December 2009 using the PACS instrument \citep{2010A&A...518L...2P} in single pointed, chop/nod spectroscopic mode. A 1D spectrum was extracted from the central 9.4" pixel of the archival fully reduced, \textit{Level 2} rebinned 
PACS data cube, following the procedure outlined in the PACS Spectroscopy Data Reduction Guide. For each spectral channel, this includes a point source correction to take 
into account the shape of the PACS spatial beam, and further rescaling by the total flux in the central 3x3 pixels to avoid missing any slightly extended disc emission. The continuum level measured through a polynomial baseline fitting at 63.18 $\mu$m 
is 16.6$\pm$1.8 Jy, in line with the 70 $\mu$m measurement (16.0$\pm$0.8 Jy) from PACS photometry \citep{2010A&A...518L.133V}. The final baseline-subtracted spectrum displayed in Fig.~\ref{figherschel}, has a channel width of 0.0023 $\mu$m and 
a spectral response profile well-approximated by a Gaussian of FWHM $\sim$0.018 $\mu$m (87.5km/s at the wavelength of the OI line). The emission line is not spectrally resolved but a clear excess is present. By fitting a Gaussian with a width that matches
the instrumental resolution we obtain the integrated line flux equal to 110$\pm$15 Jy km/s (17.4 $\pm$ 2.3 $\times 10^{-15}$ erg/s/cm$^2$), where the error bars include the 11\% flux calibration accuracy. We note that this value is consistent with that reported by \citet{Bran2016} (Table 1),
once we take into account that the authors extracted the 1D spectrum directly from the central PACS spaxel, without applying any point source correction (hence their per beam units) nor our rescaling to account for extended emission outside the central spaxel.
\end{itemize}

\begin{figure}
\centering
   \includegraphics[width=7.5cm]{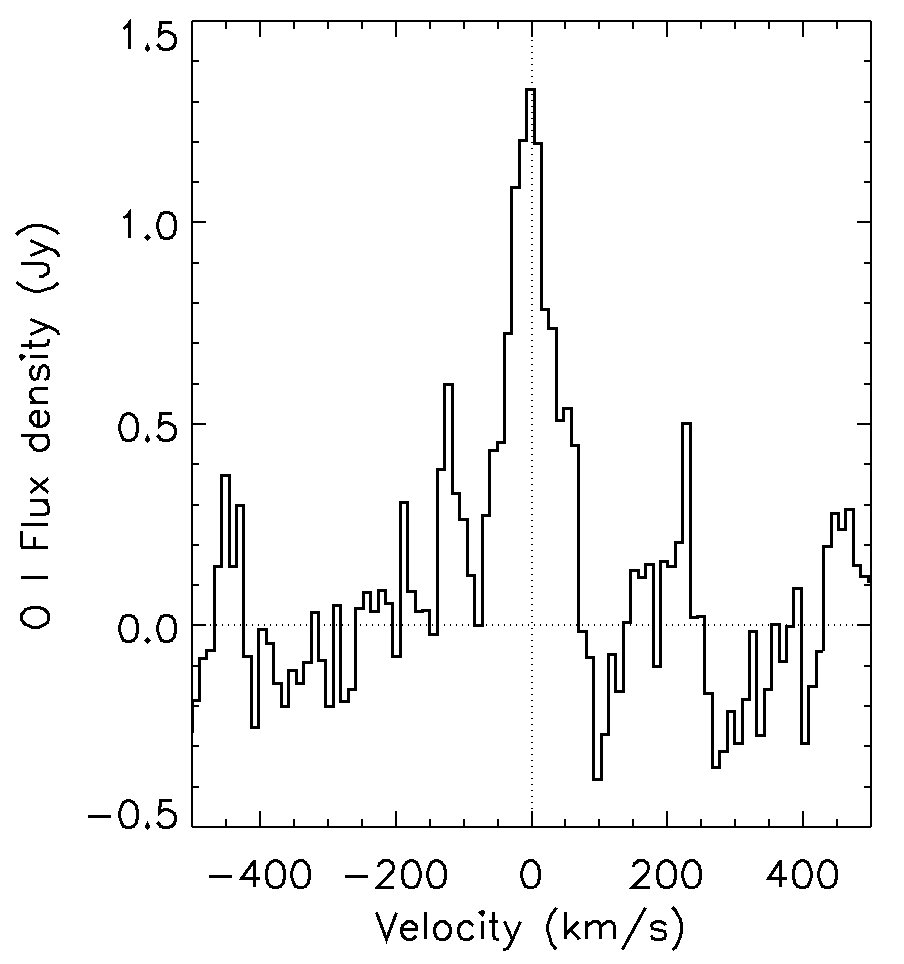}
   \caption{\label{figherschel} Herschel/PACS observation of the O I emission line profile at 63 microns centred on the star's velocity. The continuum flux equal to $\sim$ 16.6 Jy has been substracted in this plot.}
\end{figure}

\subsection{$\chi^2$ analysis to fit the C II Herschel spectrum}\label{xhi2}

\begin{figure*}
    \centering
    \begin{minipage}{.5\textwidth}
        \centering
        \includegraphics[width=0.9\linewidth]{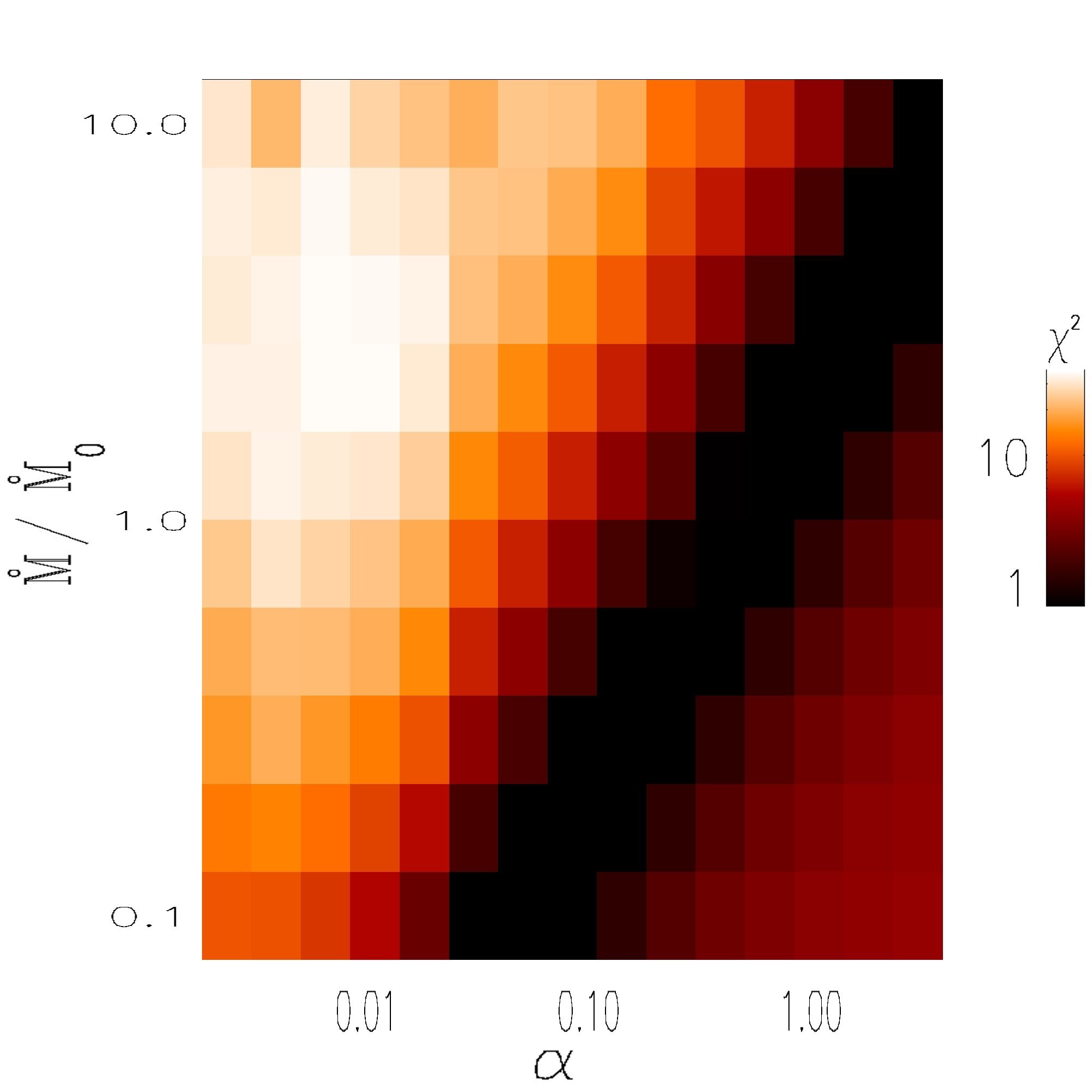}
    \end{minipage}%
    \begin{minipage}{0.5\textwidth}
        \centering
        \includegraphics[width=0.9\linewidth]{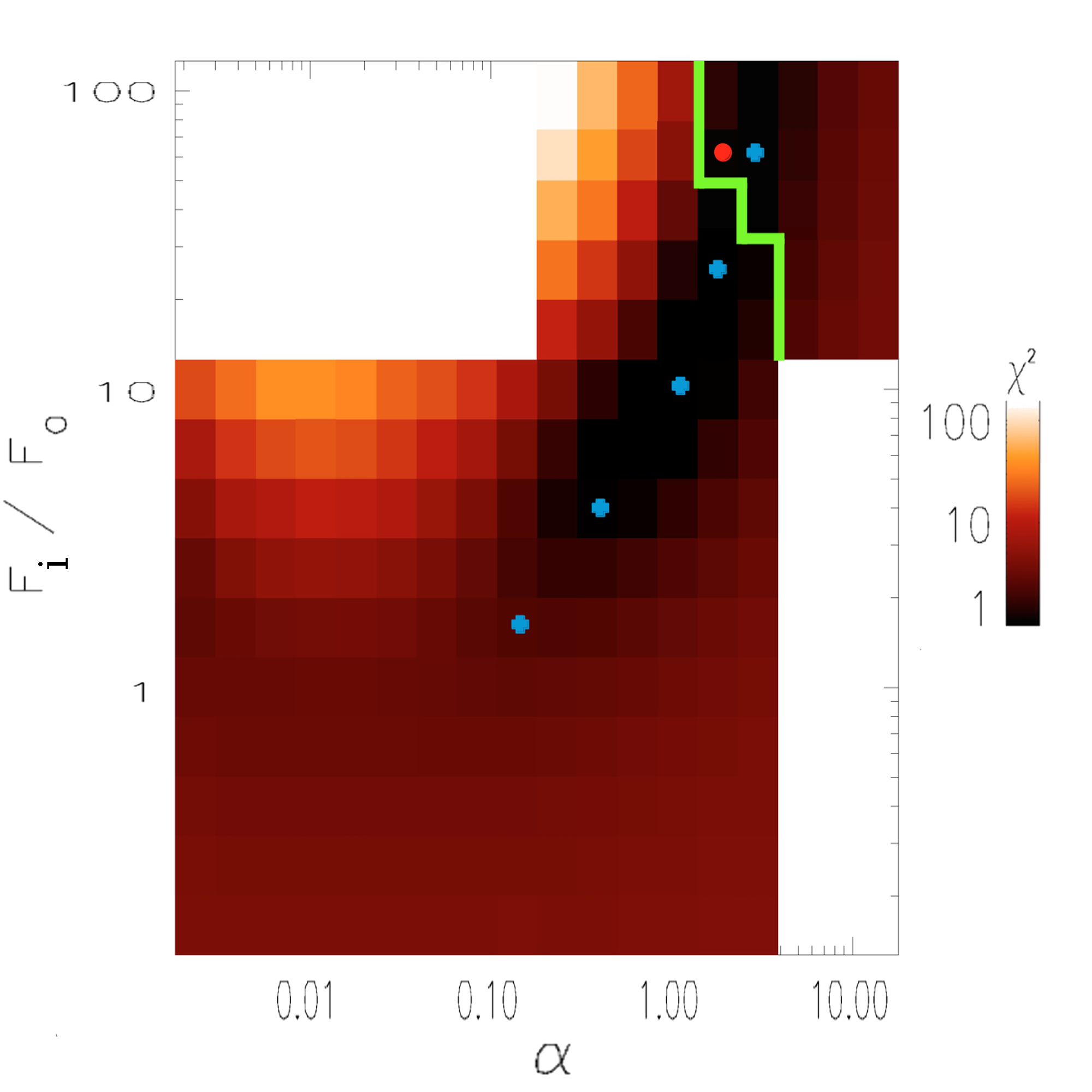}
    \end{minipage}
   \caption{\label{figchi2} Reduced $\chi^2$ map: \textit{Left}: $F_i/F_0=5$, $0.0023<\alpha<3$ and $0.1<\dot{M}/\dot{M}_0<10$. \textit{Right}: $\dot{M}/\dot{M}_0=1$, $0.0023<\alpha<13.9$ and $0.1<F/F_0<100$. Models to the right of the green line respect the C I non-detection
by APEX (see subsection \ref{CIfit}). The blue points show the different models used to compute Fig.~\ref{figionisat}. The red dot shows the best-fit model described in subsection \ref{predi}.}
\end{figure*}

Simulations that cover the parameter space of $\dot{M}$, $\alpha$ and $F_i$ were run from $t=0$ where the disc is devoid of gas until it reaches steady state. We do not present the transient evolution as the age of $\beta$ Pic is much larger than the typical viscous timescale that we find for 
our best-fit model, and so we expect the gas disc to be at steady state (unless gas production has only started very recently). We recall that in our fiducial model the carbon gas is input at $R_0$=85 AU at a rate $\dot{M}_0 = 0.1$ M$_\oplus$/Myr but given uncertainties in the CO mass 
determination, $\dot{M}$ can vary by a factor a few from $\dot{M}_0$ and is thus left as a free parameter \citep{2014Sci...343.1490D}. The three free parameters $\dot{M}$, $\alpha$ and $F_i$ are constrained by comparison with the C II observation in Fig.~\ref{figCII}.

We compare our numerically evaluated synthetic spectrum (using RADMC-3D) to the C II HIFI spectrum using a $\chi^2$ analysis. We first probe the parameter space in Fig.~\ref{figchi2} (left) for a fixed $F_i=5F_0$ and varying $\alpha$ and $\dot{M}$ with $0.0023<\alpha<3$ and $0.1<\dot{M}/\dot{M}_0<10$ (15 and 10 logarithmically spaced bins respectively).
The reduced $\chi^2_r$ for a given model was calculated as $\chi^2_r=1/N_\textrm{\tiny dof} \times \sum_i (o_i-c_i)^2/\sigma^2$, where $o_i$ is the observed flux, $c_i$ the flux given by the best-fit model for different values $i$ on the spectrum along the $x$-axis in Fig.~\ref{figCII}. For this
observation, $\sigma=6\times 10^{-16}$ erg/s/cm$^2$ and the number of degrees of freedom $N_\textrm{\tiny dof}=N-N_\textrm{\tiny fp}=20$, $N$ being the number of points used on the spectrum and $N_\textrm{\tiny fp}=3$, the number of free parameters. 
We find that there are many good fits to the data as there is a degeneracy between $\alpha$ and $\dot{M}$. This is expected, since in steady state $\Sigma \propto \dot{M}/\alpha$. 
Assuming that $\dot{M}=0.1$ M$_\oplus$/Myr within a factor two (from ALMA observations), this provides the best fits when $0.2 < \alpha < 1$ (when $F_i/F_0$ is fixed equal to 5). 

However, the best-fit depends on the UV radiation field $F_i$ impinging the disc, 
which is not well-known. To test the dependence on the IRF, we fix $\dot{M}/\dot{M}_0=1$ and we create a second $\chi^2$ map, shown in Fig.~\ref{figchi2} (right) probing models with $0.0023<\alpha<13.9$ and $F_i/F_0$ varying from 0.16 to 100. We do not explore the full parameter space
for high values of $\alpha$ and $F_i$ but restrict the study to the relevant parameters (195 simulations as seen on Fig.~\ref{figchi2} right). We find that our results are strongly dependent on the amount of UV photons impinging on the $\beta$ Pic disc. 
We note that the best fits are for $3<F/F_0<100$. The excess of UV flux compared to the standard IRF could be coming from the star or the environment close to the star (see discussion). 

\begin{figure}
\centering
   \includegraphics[width=7.5cm]{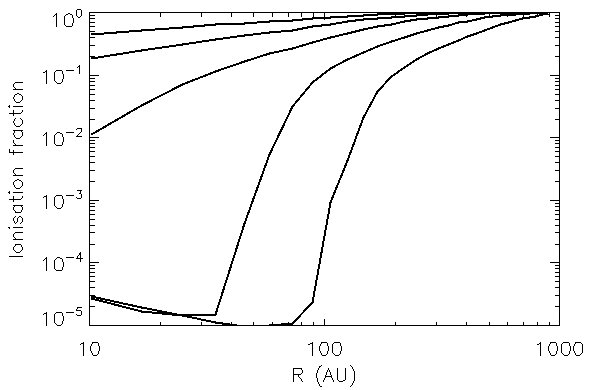}
   \caption{\label{figionisat} Carbon ionisation fraction for different $\alpha$ and impinging UV flux for models represented with a blue dot on Fig.~\ref{figchi2} (right) for which the parameters equal [$\alpha$,$F_i/F_0$]=[(0.14,1.6);(0.39,4);(1.1,10);(1.8,25);(3,63)] from bottom to top lines.}
\end{figure}

To explain the location of the best-fit models on Fig.~\ref{figchi2} (right), note that we are keeping $\dot{M}$ constant but changing $\alpha$ and so the density is lower for higher $\alpha$. Despite the lower density, it can still be possible to fit the C II line if the ionisation fraction is increased, which can be achieved by increasing the UV radiation. This is indeed
what is happenning in the models as shown on Fig.~\ref{figionisat}. Very large values of $\alpha$ cannot fit the C II line because the ionisation fraction cannot go above one. The lack of good fit for $F_i/F_0 < 4$ arises for a different reason, which is that the higher densities
cause the disc to become optically thick to FUV photons in the inner regions. The absence of C II at high velocities then also changes the shape of the line, which otherwise fits very well (see Fig.~\ref{figCII}).


We emphasise that these two maps provide sufficient information to consider the whole parameter space of $\dot{M}/\dot{M}_0$, $F_i/F_0$, and $\alpha$ due to the degeneracy between $\alpha$ and $\dot{M}$, i.e, we do not need to create a 3D $\chi^2$ map.
In steady state, a decrease in $\alpha$ is the same as an increase in $\dot{M}$ so that one could create a new map for $\dot{M}/\dot{M}_0=2$ (the upper limit derived from the ALMA CO observation) by shifting the map to the left in Fig.~\ref{figchi2} (right) by a factor 2 in $\alpha$.
Note that $\dot{M}$ could be even greater if some other molecules such as CH$_4$ or CH$_3$OH are also present so that the $\alpha$ value required would get smaller (see the discussion).

We are left with a range of models that still fit the C II observation. We need to feed in more observations to our model to constrain the best-fit. We see in the two next subsections how the APEX C I non-detection and the Herschel/PACS O I spectrum can help to constrain the best-fit. 

\subsection{How to also fit the APEX C I non-detection?}\label{CIfit}

On Fig.~\ref{figchi2} (right), one can see that the best fits for the C II line still allow two directions, one towards a smaller impinging UV flux and smaller $\alpha$ or a greater UV flux and higher $\alpha$. 
As discussed in subsection \ref{xhi2}, the best fits are achieved by changing the ionisation fraction, so while C II remains fixed on the black part of Fig.~\ref{figchi2} (right), the higher $\alpha$, $F_i/F_0$ models have lower overall surface density and higher ionisation and so less C I.

For each best-fit model in Fig.~\ref{figchi2} (right), we compute the C I line and compare the total integrated flux to the limit imposed by the non-detection (14 Jy km/s). The non-detection implies that only models to the right of the green line in Fig.~\ref{figchi2} (right) are allowed.
We find that the models that best fit the C I non-detection are for lower C I densities, i.e. for higher UV flux impinging on the disc and greater $\alpha$. We are left with only 4 models that fit both the C I non-detection
and C II spectrum. We choose the best-fit as being the one with the smallest $\alpha$ value that respects the conditions. The chosen best-fit is shown as a red dot on Fig.~\ref{figchi2} (right).
 We thus find that our best-fit parameters are $\dot{M}/\dot{M}_0=1$, $\alpha=1.5$, $F_i/F_0=60$.


On Fig.~\ref{figCII} and \ref{figCI} we overplot our best-fit model with the Herschel C II line and the APEX non-detection respectively.
Our predicted emission lines are in black and observations in red. 

For C II, the total emission line flux worked out by our model for this best-fit is 365 Jy km/s, which is within 2\% of the observed value 372 Jy km/s. The best fit not only takes the total integrated flux into consideration but also the shape of the emission line and 
its peak value through our $\chi^2$ analysis. 

As for C I, we find a total flux of 11 Jy km/s where the APEX non-detection imposes that the 3$\sigma$ value must be smaller than 14 Jy km/s. \citet{2014A&A...563A..66C} predicted a C I total flux of 55 Jy km/s (their multiple ring model), which would have been observed by APEX if true. 
This non-detection and best-fit will enable a much more constrained prediction for ALMA observations of C I (see subsection \ref{predi} for details). 

The O I spectrum can now be used to check the consistency of our results or even to derive new quantities as explained below.

\subsection{How to also fit the Herschel O I spectrum and derive the hydrogen content?}\label{Hline}

\begin{figure}
\centering
   \includegraphics[width=7.5cm]{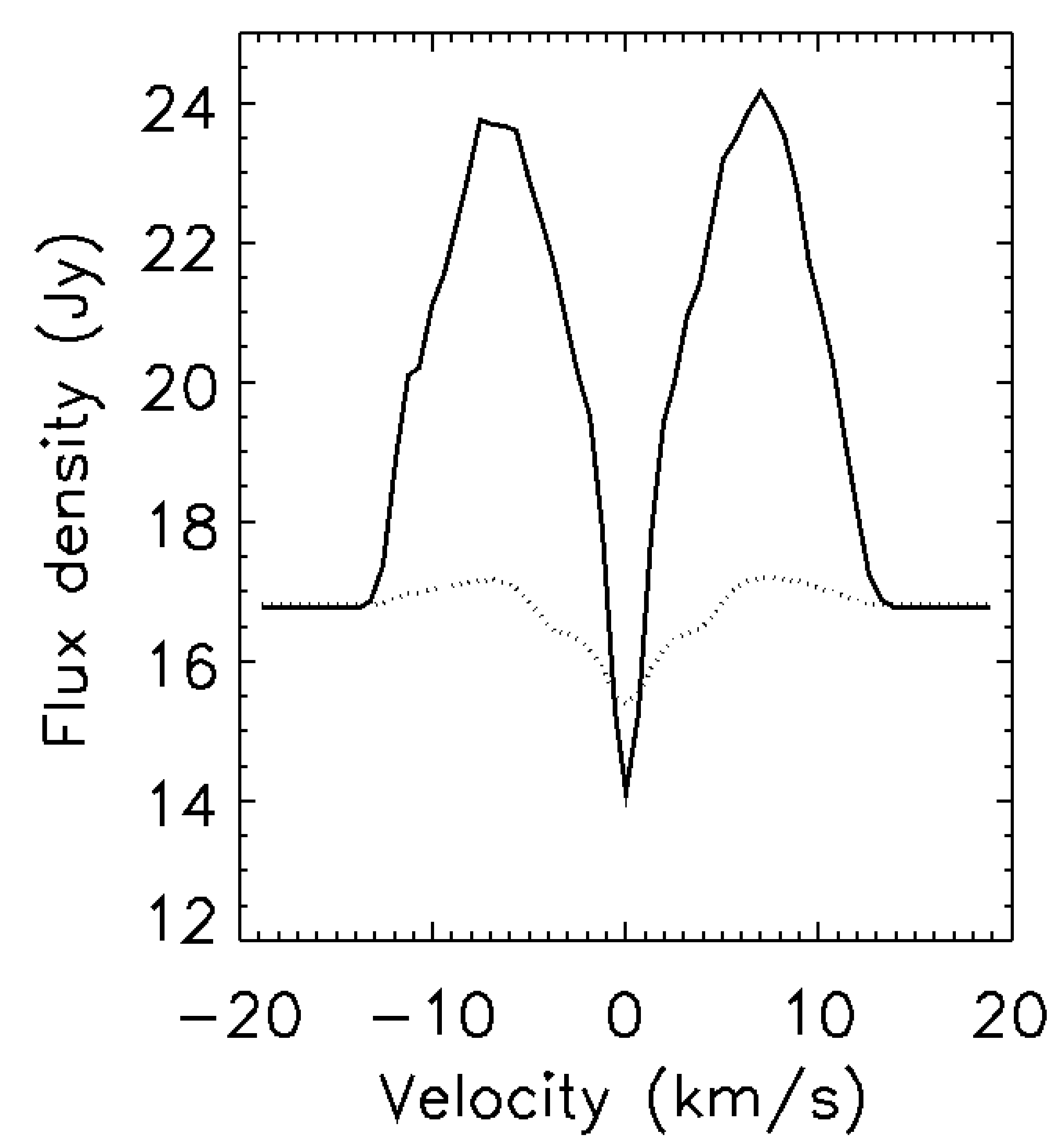}
   \caption{\label{figOIline} O I emission line profile at 63 microns for our best-fit model. The dotted line is when only electrons are considered as collisional partners. The solid line is when a water component with H/C $\sim$ 3 is added. In the last case, it fits the total integrated flux from the observed
line shown on Fig.~\ref{figherschel}. The continuum flux is not substracted in this image contrary to Fig.~\ref{figherschel} and equals $\sim$ 16.6 Jy.}
\end{figure}

Fig.~\ref{figOIline} shows the predicted O I emission line at 63 microns for our best-fit model. The critical density of O I is on the order of $10^{4}$ cm$^{-3}$, which is much higher than the electron density derived with our model everywhere in the gas disc. 
Hence, LIME was used to make a full NLTE calculation of the O I line.
The O I line observed by Herschel is not spectrally resolved (nor spatially) and only the total emission flux predicted by our model should be compared, not the shape of the line. As noted in Table \ref{tab1}, the total integrated flux observed is equal to 110 $\pm$ 15 Jy km/s. 

While the density suggests that NLTE calculations are required, this calculation is complicated by the lack of collision coefficients for collisions of O with anything but e$^-$ and H, whereas there is no H in the model. For our best-fit model, we first try to calculate the O I line in NLTE with LIME using electrons as the only colliders with oxygen. We use the electron density predicted for our best-fit model as well as the dust-to-gas mass ratio (see Fig.~\ref{figdens}a,f). 
The NLTE calculation with LIME gives a total integrated flux equal to -2 Jy km/s (absorption), which is lower than the observed value. The resulting line is shown on Fig.~\ref{figOIline} (dotted). The line shows both emission at large velocities and absorption in the centre of the line. This can be understood
looking at Fig.~\ref{figdens}a as most of the O I emission comes from the inner part of the system and the medium is optically thick.
We also check that the continuum value derived is close to the observed value (see subsection \ref{obsc}). 

However, we can expect that the addition of more colliders in the calculation will be able to fit the observation, since an LTE calculation gives a flux 22 times that observed. Since collision rates with C I, C II and O I are unknown, here we ask the question what level of H
is required to fit the observation. We consider this hydrogen component within the context of a model in which it arises from water released in the same process as the release of CO. H$_2$O photodissociates even faster than CO \citep[e.g.][]{2015MNRAS.447.3936M} and so this would result in hydrogen and some more oxygen being released into the gas disc. These atomic species have an ionisation potential of 13.6eV and so stay neutral. H I is not pushed by radiation pressure and should have a similar evolution to O I. With the help of the Herschel/PACS O I line, one can then
assess the amount of H$_2$O released with CO needed to fit the line. 
We add the hydrogen into LIME as a new collisional partner along with the electrons and quantify the amount of H$_2$O required
to fit the observation. In the process, we also add the right amount of oxygen released from water to the amount of oxygen released from CO. We find that fixing a ratio H/C $\sim$ 3 gives the best fit to the data, which translates into an H$_2$O/CO ratio of $\sim$ 1.5. The resulting plot is shown in Fig.~\ref{figOIline} (solid) and the total flux found is equal to 110 Jy km/s (i.e close to the observed value). Although the line shown in Fig.~\ref{figOIline} cannot be observed with currently available instruments, future missions such as SPICA are expected to have a higher resolution and sensitivity than Herschel to measure the O I and C II lines \citep{2009ExA....23..193S} from which our model can be used to make predictions.
Assuming this scenario, one should add the oxygen coming from water to the original oxygen number density coming from CO. Overall the oxygen number density must be multiplied by a factor $\sim$ 2.5 to extract the final amount of oxygen. This translates as O/C $\sim$ 2.5 and O/H $\sim$ 1 (see Table~\ref{tab4}). Using the total CO mass derived by \citet{2014Sci...343.1490D} and the photodissociation timescales for H$_2$O and CO, one finds that the total H$_2$O mass in the gas phase is $\sim$ $2 \times 10^{-9}$ M$_\oplus$.
This does not change our other results as we checked that the temperature (which fixes the viscosity) does not vary assuming this new oxygen (and hydrogen) content.

If this model is correct in that the extra colliders needed to explain the O I line come from H$_2$O, the gas disc is depleted in hydrogen compared to the hydrogen component that would be expected from comets with Solar System-like compositions, where H$_2$O/CO varies from 3 to 250 \citep{2011ARA&A..49..471M}. 
Thus, we deduce that a typical Solar System-like comet composition can be ruled out for $\beta$ Pic. Even if one tries to go towards higher UV flux on Fig.~\ref{figchi2} (right), the best fits 
(in black, above the green line) do not have a different amount of oxygen as $\alpha$ keeps constant
when the UV flux increases (and oxygen stays neutral). Moreover, adding more colliders with oxygen, such as C I, C II or O I would only lower the total amount of hydrogen (i.e., H$_2$O in the parent debris) required to fit the OI line. 

\subsection{First measurement of $\alpha$ in a debris disc}\label{firstm}

As explained in subsection \ref{xhi2}, in steady state, a decrease in $\alpha$ is the same as an increase in $\dot{M}$. Thus, if $\dot{M}/\dot{M}_0=2$, the black part of Fig.~\ref{figchi2} (right) would shift to the left by a factor 2 in $\alpha$.
$\dot{M}_0$ is known within a factor 2 from the ALMA observation but some carbon could come from some other molecules such as CH$_4$ or CH$_3$OH although these are thought to be only a small percentage of the CO in comets. However, $\dot{M}$ could vary
by another factor $\sim$ 1.5 assuming an extreme comet composition. For a fixed $F_i/F_0$, this provides a way to constrain $\alpha$, as a factor 3 higher $\dot{M}$ would have the same effect as a factor 3 smaller $\alpha$. Hence, for the highest
mass input rate, $\alpha$ could go as low as 0.5 and still agree with observations. $F_i/F_0$ could also be lower if the C I spatial distribution is affected by the presence of planets. Overall, taking into account the observational uncertainties and the inherent model uncertainties,
 we estimate that $\alpha$ needs to be greater than $\sim$ 0.1 in $\beta$ Pic to explain all the observations.

\subsection{Details of Best-Fit Model}\label{predi}

\begin{table}
    \caption{Parameters of the best-fit model.}
\begin{center}
\begin{tabular}{|l|c|}
  \hline \hline
  $\alpha$ & 1.5  \\
  $\dot{M}$ & 0.1 M$_\oplus$/Myr  \\
  $F_i/F_0$ & 60 \\
  Star & A6V \\
  Dust & See Eq.~\ref{taudust}\\
  \hline
\label{tab2}
\end{tabular}
\end{center}
\end{table}

\begin{figure*}
\centering
   \includegraphics[width=17.5cm]{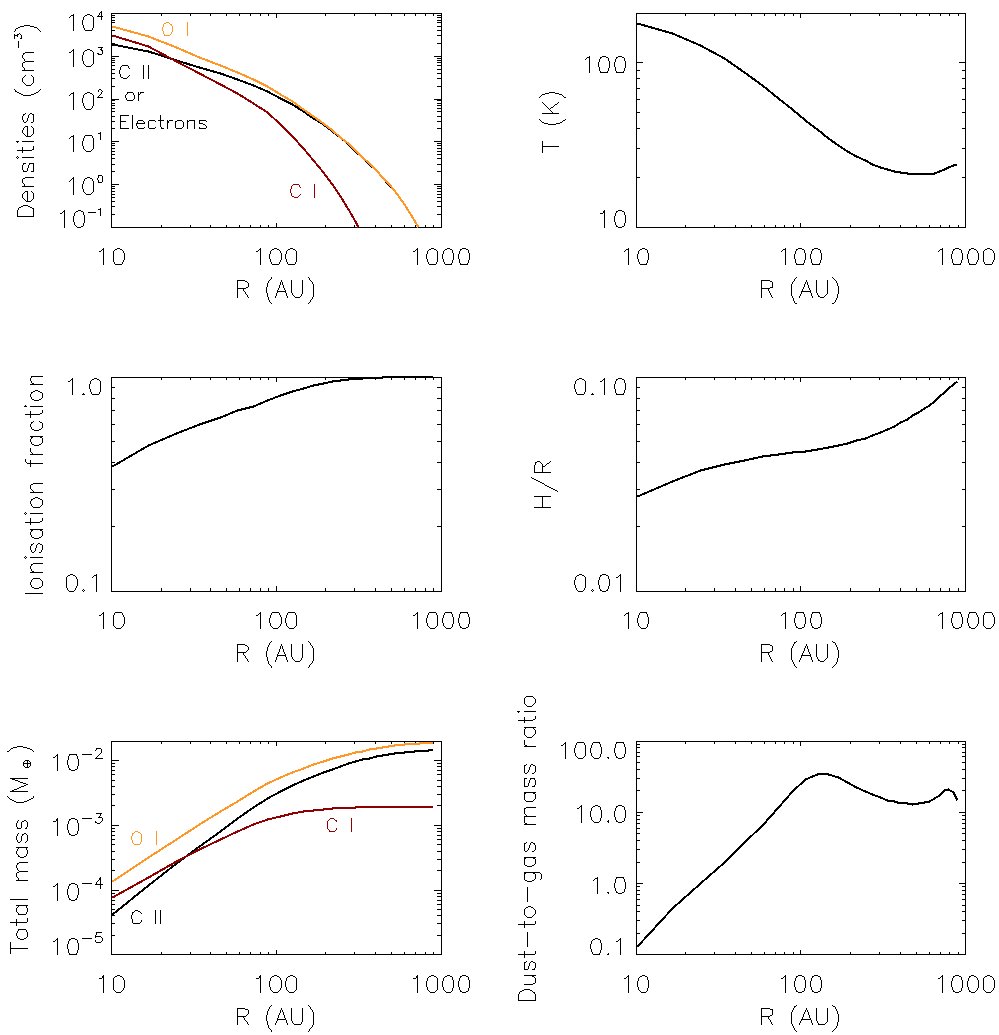}
   \caption{\label{figdens} From top to bottom and left to right, a) Densities of C I (red), C II or electrons (black) and O I (yellow); b) Temperature as a function of $R$; c) Ionisation fraction as a function of $R$; d) H over R as a function of $R$; e) Cumulative mass of C I (red) and C II (black) and O I (yellow) as a function of $R$; f) Dust-to-gas ratio as a function of $R$.}
\end{figure*}

Fig.~\ref{figdens} provides a summary of the structure of the gas disc predicted by our model in the case without extra oxygen coming from water. The best fit parameters are $\alpha=1.5$, $F_i/F_0=60$ and an injection rate equal to $\dot{M}_0 = 0.1$ M$_\oplus$/Myr (see Table~\ref{tab2}), which is the value given by the ALMA CO observation \citep{2014Sci...343.1490D}.

Fig.~\ref{figdens}a gives the spatial profile of C I in red, C II in black as well as the oxygen density in yellow. The electron density is superimposed on the C II density as all electrons come from the photoionisation of C I in our model. The C I density is smaller than that of C II when $R>20$AU due to the increasing ionisation fraction with $R$. The O I density is the sum of C I and C II as
almost no oxygen gets ionised and C/O=1. In the case with extra oxygen coming from water, the O I density shown on Fig.~\ref{figdens}a should be multiplied by $\sim$ 2.5.

One of the specificities of the observed C II emission line is that the velocity gradient is very steep and does not show broad wings as would be expected from a naive accretion disc profile. 
Instead, the C II density $<80$ AU scales as $R^{-1.15}$ (i.e $\Sigma \propto nH \propto R^{-0.15}$) as can be seen on Fig.~\ref{figdens}a, which is shallower than the usual $\Sigma \propto R^{-1}$ assumed in other studies \citep[e.g.][]{2014A&A...563A..66C}. 
This shallow profile arises because of the decrease of ionisation and increase in temperature towards smaller radii. The reduced ionisation fraction in the inner regions (see Fig.~\ref{figdens}c) is a consequence of the gas disc becoming optically thicker to 
FUV radiation as the total C I density increases (since C I absorbs the UV flux before it reaches the midplane) which in turn reduces the amount 
of C II in the inner regions. Thus, while \citet{2014A&A...563A..66C} concluded that the shape of the C II line is inconsistent with an accretion disc profile, we find that the profile from an accretion disc is a good fit to that observed.

The C II densities obtained with our best fit model can be compared to those derived from a previous simpler model. Fitting the C II HIFI emission line with a series 
of four rings, \citet{2014A&A...563A..66C} obtained a best fit to the total carbon mid-plane density (see their Table 3 and the large error bars associated). They obtained very high ionisation fractions (higher than 0.5) similar to that derived from our
model. The most reliable comparison is within their 30-150 AU annulus, where their error bars are smaller, where they find a density of $\sim 100$ cm$^{-3}$, which is in good agreement with our best-fit where the density
 is equal to $\sim$ 110 cm$^{-3}$ at $\sim$ 100AU. Our model extends these values to a range of densities as a function of radius, giving an electron and a C II density varying as $R^{-1.15}$ when $R<100$ AU. It then falls off quicker to reach 10 cm$^{-3}$ at R $\sim$ 230 AU.

Fig.~\ref{figdens}b gives the temperature profile expected for the gas. From 10 to 20 AU, the temperature scales as $R^{-1/3}$ and then falls off as $R^{-0.8}$ up to 200AU before reaching a plateau at $\sim$ 20K.
This profile can be explained by looking at the ionisation profile, which shows also three different regimes when $<$20AU, between 20-200AU and when $>200$ AU. 
The decrease in temperature is mainly due to the ionisation fraction becoming higher towards the outer region so that cooling by C II increases. In the inner part, another complication comes into play since the increased temperature means that O I also contributes to the cooling reaching up to 40\% of the total cooling
rate.

The corresponding scale-height is given as a function of $R$ on Fig.~\ref{figdens}d. Here also 3 regimes can be distinguished. Between 20 and 200AU, $H/R$ is shallow and scales $\propto R^{0.1}$. A similar almost linear variation of $H$
with $R$ was observed for Fe I with VLT and gives an indication that the gas could be well-mixed \citep{2012A&A...544A.134N}. However, $H/R$ observed for Fe I is on the order of 0.2, which is higher than expected by our model. This difference
might imply that the temperature of Fe is totally decoupled from the main gas disc. In the inner region, as the temperature drops, $H$ gets smaller, and so does $H/R$ reaching 0.028 at 10AU. Unfortunately, the regions inwards of 40AU of the spatially resolved observation 
of Fe I are very noisy due to the PSF substraction and a confident $H/R$ value could not be extracted here.

The electron density in $\beta$ Pic is represented by the black line on Fig.~\ref{figdens}a, and follows the C II density, since in the model all electrons are found to come from the photoionisation of carbon. Thus, the electron density falls off as $R^{-1.15}$.
An alternative independent method to derive the electron density involves using the ratio of CO J=2-1 \citep{2016MNRAS} and CO J=3-2 line fluxes, since this ratio is set by the density of colliders. 
If these are assumed to be electrons, the radial dependence of the electron density from our model agrees with that derived using this alternative method within the error bars of the observations 
\citep{2016MNRAS}. However the absolute electron density predicted by these two methods differs by a factor $\sim$ 2, which could suggest that electrons are not the only colliders with CO, and that C II and O I for instance could have an impact on the overall CO excitation.

The cumulative masses of C I, C II and O I over the whole disc are plotted on Fig.~\ref{figdens}e. The total mass observed for a given field of view (FOV) or FWHM can be obtained by looking at the value at which $R$ roughly equals the maximum FOV, which can be useful to compare to
masses derived from observations. The total mass over the whole gas disc of C I is $2 \times 10^{-3}$ M$_\oplus$, $1.3 \times 10^{-2}$ M$_\oplus$ for C II, and $2 \times 10^{-2}$ M$_\oplus$ for O I (see Table~\ref{tab3}). This  should not be compared directly to the millimetre dust mass 
of $6 \times 10^{-2}$ M$_\oplus$ \citep{2009A&A...508.1057N}, since the bulk of C I is in the inner region, and the dust mass is located mainly between 50-130AU. The dependence of the dust-to-gas mass ratio is plotted on Fig.~\ref{figdens}f, where we assumed a standard size distribution $q=-3.5$, from $s_{\rm min}$=5$\mu$m 
(blow-out size) to $s_{\rm max}=1$mm and the optical depth profile described in subsection \ref{dustdisc}.

\begin{figure}
\centering
   \includegraphics[width=7.5cm]{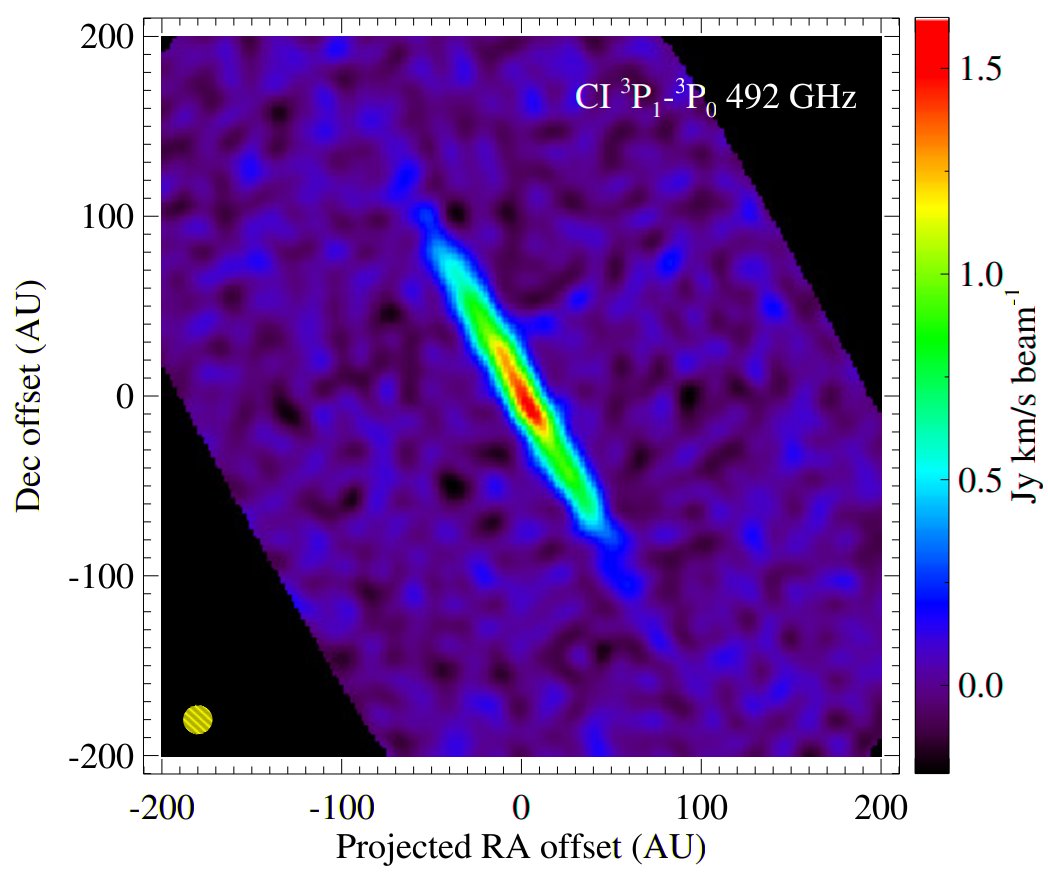}
   \caption{\label{figalmaC1} Synthetised ALMA (Cycle 3 in compact configuration) moment-0 image for C I emission at 609 microns for our best-fit model. The beam size is 0.84''x0.83'' and the total flux equals 12 Jy km/s (see the text for details).}
\end{figure}

We used the model to make predictions for ALMA observations of C I. The setup is analogous to that employed for the ALMA prediction in \citet{2014A&A...563A..66C}, 
namely 1.24h on-source time spread between 3 pointings across the disc midplane (6" apart), in compact configuration, and standard weather conditions for 492.16 GHz observations (0.472mm precipitable water vapor).
We used the \textit{simobserve} task within the CASA software version 4.5.0 \citep{2007ASPC..376..127M} to produce the visibility dataset, then used the CLEAN algorithm (with natural weighting of the visibilities) to 
produce a synthetic data cube with a final channel width of 0.63 km/s and synthesised beam of 0.84''x0.83''. Fig.~\ref{figalmaC1} shows the resulting spectrally integrated (i.e. moment-0) image of the CI $^3$P$_1$-$^3$P$_0$ emission from the disc
for which the recovered integrated line flux is only slightly lower than predicted by our model (12 as opposed to 13.5 Jy km/s). Comparison with the observed total flux and radial/vertical brightness distribution will provide a strong test of the model predictions.

\begin{figure}
\centering
   \includegraphics[width=7.5cm]{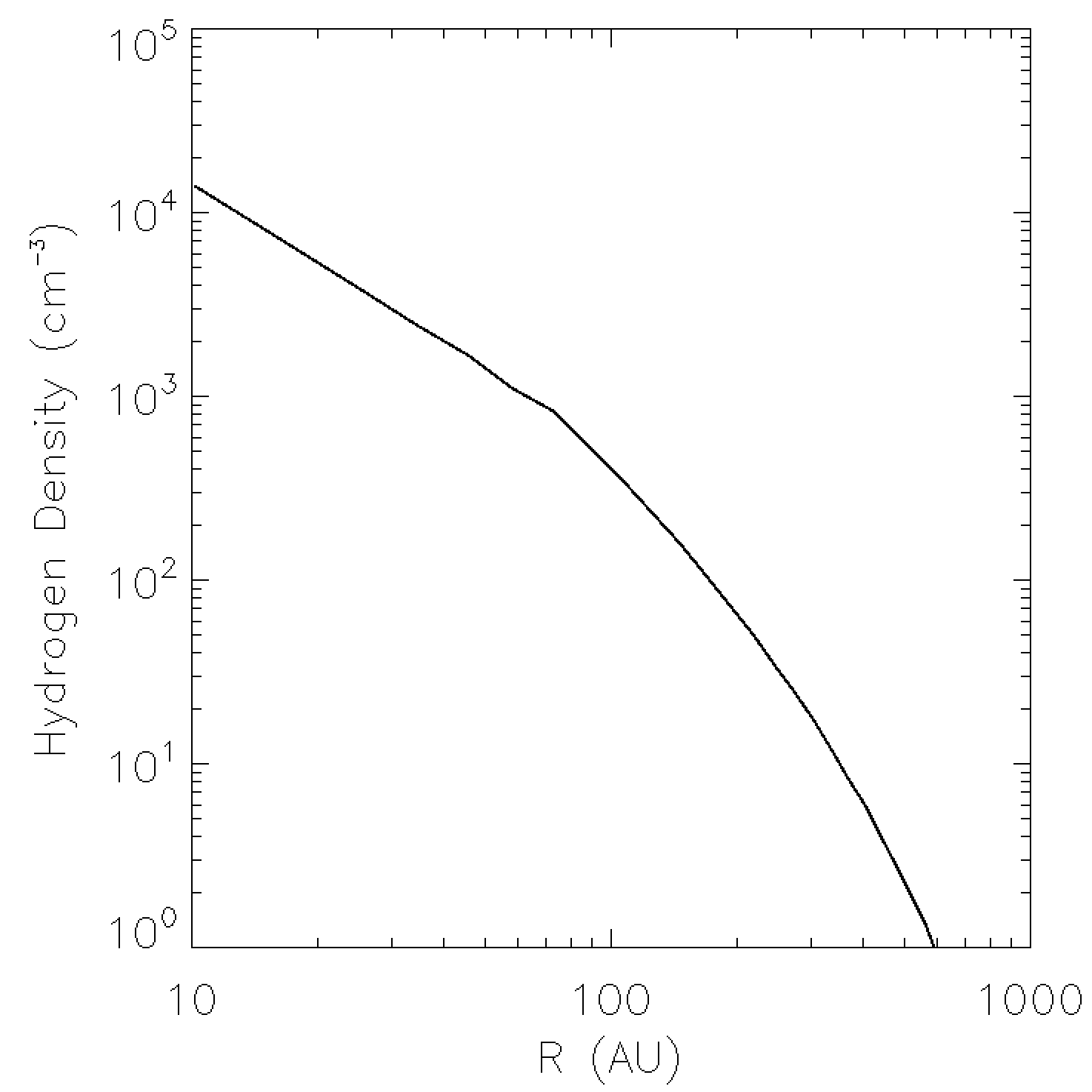}
   \caption{\label{figHIline} H I number density predicted for our best-fit model. The total H I column density along the line-of-sight is $\sim 3\times 10^{18}$ cm$^{-2}$.}
\end{figure}

In subsection \ref{Hline}, we computed that to fit the O I Herschel line requires an H/C ratio of $\sim$ 3. Thus, we can make predictions for the total amount of hydrogen within the system.
Fig.~\ref{figHIline} shows the derived density for hydrogen for our best-fit model, for which we derive the total H I column density along the line-of-sight is $\sim 3\times 10^{18}$ cm$^{-2}$ (which may be observable as hydrogen absorption lines in UV). We also
derive the total H I mass to be $3.1 \times 10^{-3}$ M$_\oplus$.

Another model prediction is that there should still be accretion in $\beta$ Pic at a rate which is roughly equal to the CO mass input rate evaluated to be $\sim 1.4\times 10^{18}$ kg/yr (within a factor 2). This value could be lower if $\beta$ Pic b accretes some of the gas before it reaches the star. 
However, this accretion rate assumes that only CO contributes to the production of C and O, and so the accretion rate could also be higher if some other molecules are produced at the same time CO is released. Indeed, if H$_2$O is also created, as inferred in subsection \ref{Hline} from the O I line, the accretion rate would increase to $\sim 2.55\times 10^{18}$ kg/yr. 
Usually, mid-A to mid-B type stars are known to be X-ray dark but $\beta$ Pic shows some weak X-ray emission that could be coming from the thermal emission of a cool corona or from some remnant accretion onto the star \citep{2012ApJ...750...78G}.
\citet{2005A&A...440..727H} estimates that an accretion rate between $2\times 10^{18}$ kg/yr and $2\times 10^{20}$ kg/yr can explain the X-ray observation, which fits our model predictions.

\begin{table}
    \caption{Total gas mass for different species.}
\begin{center}
\begin{tabular}{|l|c|c|}
  \hline \hline
  Element & Best-fit & Best-fit with extra hydrogen \\
  \hline
  C I & $2 \times 10^{-3}$ M$_\oplus$ & $2 \times 10^{-3}$ M$_\oplus$ \\
  C II & $1.3 \times 10^{-2}$ M$_\oplus$ & $1.3 \times 10^{-2}$ M$_\oplus$ \\
  O I & $2 \times 10^{-2}$ M$_\oplus$ & $5 \times 10^{-2}$ M$_\oplus$\\
  H I & $\sim$ 0 & $3.1 \times 10^{-3}$ M$_\oplus$\\
  H$_2$O & $\sim$ 0 & $2 \times 10^{-9}$ M$_\oplus$\\
  \hline
\label{tab3}
\end{tabular}
\end{center}
\end{table}

\begin{table}
    \caption{Gas number density ratio when assuming extra hydrogen in the disc.}
\begin{center}
\begin{tabular}{|l|l|c|}
  \hline \hline
  & H/C & $\sim$ 3 \\
  In the gas disc & O/C & $\sim$ 2.5  \\
  & O/H & $\sim$ 1  \\
  \hline 
  In planetesimals &H$_2$O/CO & 1.5 \\
  \hline
\label{tab4}
\end{tabular}
\end{center}
\end{table}

\subsection{Dust in $\beta$ Pic}\label{dustbp}

We use the dust density profile described in subsection \ref{dustdisc} to compute the effect of photoelectric heating on the temperature within the disc using Cloudy. When dust is added, we see no effects on the density profile nor on the emission line predictions. 

This can be explained in the following way. Using the definitions of $\Gamma_\mathrm{PE}$ and  $\Gamma_\mathrm{ionC}$ (photoelectric and photoionisation heating rates) and the fact that the photoelectric charging current per unit area on each dust grain
 is equal to the thermal electron collection current \citep{2010ApJ...720..923Z}, one finds that to have 
$\Gamma_\mathrm{PE} > \Gamma_\mathrm{ionC}$ requires that

\begin{equation}
\label{dustcond}
 \tau_d > \tau_\mathrm{crit} = \frac{H n_{C_\mathrm{I}} \langle E_\mathrm{ion} \rangle \sqrt{2 \pi k_B T m_e}}{4 t_\mathrm{ionC}  \langle E_\mathrm{dust} \rangle s_e n_e e \phi},
\end{equation}

\noindent where $n_{C_\mathrm{I}}$ and $n_e$ are the C I and electron densities respectively, $m_e$ is the electron mass, $s_e \sim 0.5$ is the electron sticking coefficient, $\langle E_\mathrm{dust} \rangle$ is the mean energy imparted by the electrons to the gas (about a few eV),
$\langle E_\mathrm{ion} \rangle$ is the mean initial energy of the ejected electron after a photoionisation and $t_\mathrm{ionC}=n_{C_\mathrm{I}}/R_\mathrm{ionC}$ is the ionisation timescale.
We also assumed that $e\phi \gg k_B T$, $\phi$ being the charging potential of the grain, as $e \phi$ is generally a few $k_B T_\star$. This formula gives a good order of magnitude for the amount of dust required for the photoelectric 
heating to be more efficient than the photoionisation heating. 

All the values within Eq.~\ref{dustcond} are given by our model so that 
for a given $R$, one can work out the effect of photoelectric heating on the disc. At $R \sim 120$ AU, where the dust density is maximum, $H \sim 5$ AU,
$n_{C_\mathrm{I}} \sim 10$ cm$^{-3}$, $n_e \sim 80$ cm$^{-3}$, $T \sim 40$ K, $t_\mathrm{ionC} \sim 2$ years and $e \phi \sim 8000 k_B$.  $\langle E_\mathrm{dust} \rangle$, the energy gained for each
electron ejected from the dust grains is about a factor 2 greater than that ejected from C I as the ionisation potential is higher than that needed to be ejected from the grain (i.e $e\phi+W$, where $W$ is the work function of the grains). The result is that $\tau_d > 10^{-2}$ is necessary
for the photoelectric effect to dominate over the photoionisation of carbon. This value is never reached in $\beta$ Pic, and in debris discs in general, explaining why the inclusion of photoelectric effect has no effects on our results.

Previous models had concluded that photoelectric heating was dominant in most of the $\beta$ Pic disc \citep{2010ApJ...720..923Z}. However, the density of carbon used in such models was a lot lower than now known to be the case. Had we used the same densities 
as assumed in \citet{2010ApJ...720..923Z}, we would also have concluded that photoelectric heating dominates photoionisation heating (as seen in their Fig.~1).

\begin{figure}
   \centering
   \includegraphics[width=8.5cm]{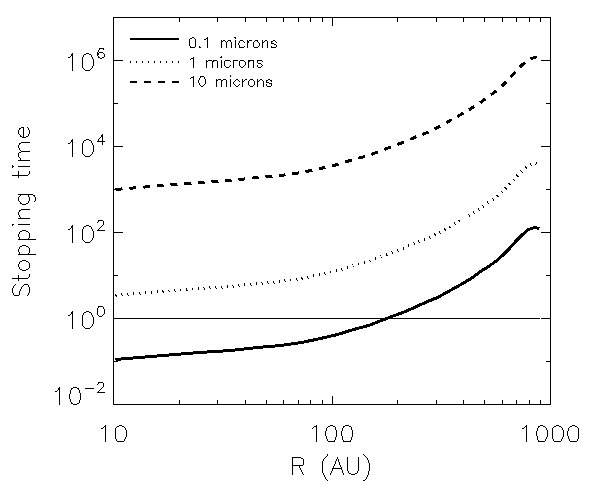}
   \caption{\label{figstop} Dimensionless stopping time versus $R$ for 3 different grain sizes 0.1 (solid), 1 (dotted) and 10 microns (dashed). The thin solid line shows when the stopping time equals 1.}
\end{figure}

Although there is no thermal effect from the dust, gas could have a dynamical effect on dust grains in these inner regions as the gas density is rather high. To quantify this, we worked out the dimensionless stopping time for our best-fit model.
Fig.~\ref{figstop} shows the stopping time as a function of $R$ for three different grain sizes
0.1 (solid), 1 (dotted) and 10 microns (dashed). In $\beta$ Pic, the blow-out size due to radiation pressure is close to 5 microns. For the case with extra oxygen coming from water, the stopping times shown on Fig.~\ref{figstop} should be divided by a factor $\sim$ 3. One can see that the grains need to be very small for the gas to brake them. Submicron grains feel a drag from
$\sim$ 200 AU inwards. This means that the grains produced in the main belt that are expected to be on unbound trajectories (i.e those below the blow-out limit) can instead become coupled to the gas and be affected by gas drag so that they drift inward instead of being ejected. However, bound grains just above the blow-out limit should not be significantly affected by gas drag even in the inner region. 

The latest observations by GPI \citep{2015ApJ...811...18M} show a dust density profile that falls off as $R^{-0.85}$, with small grains as close in as 23AU. Since the ALMA
observations show that the main belt is beyond $\sim$ 50AU, these small grains cannot be produced in-situ. The high gas density we predict and the resulting drag on sub-micron grains may help to explain this observation. This may also explain mid-IR observations which show material within 20AU \citep{1997A&A...327.1123P}.


\subsection{$\alpha$ model explained with MRI?}

\begin{figure*}
   \centering
   \includegraphics[width=17.5cm]{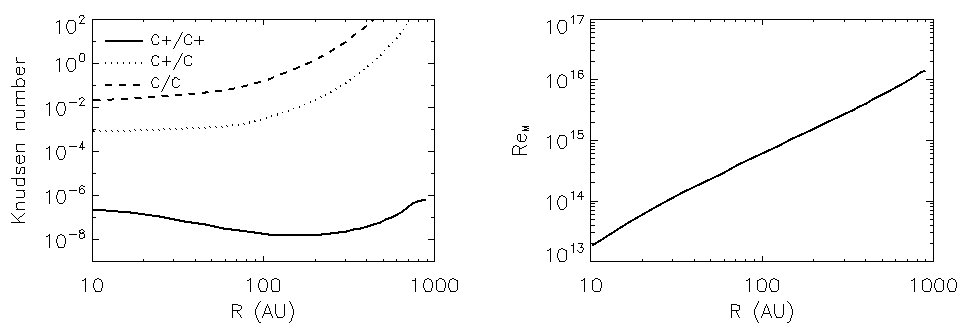}
   \caption{\label{Knudsenplot} \textit{Left:} Knudsen Kn number in $\beta$ Pictoris for C$^+$/C$^+$ collisions (solid line), C$^+$/C collisions (dashed), C/C collisions (dotted). \textit{Right:} Magnetic Reynolds number $Re_M$ in $\beta$ Pictoris (see Eq.~\ref{Rem}).}
\end{figure*}

In Figure \ref{Knudsenplot} (left), we plot the Knudsen number Kn=$\lambda_{i,j}/H$ for our best-fit disc, where $\lambda_{i,j}$ is the mean free path of element $i$ with $j$. We see that throughout most of the disc (where densities are higher than 10 cm$^{-3}$), 
taking the mean free path between C$^+$/C$^+$ (solid line), C$^+$/C (dotted) or C/C (dashed), Kn $ < 1$, and thus we conclude that modelling this disc as a fluid should give an adequate description.

The value of the viscosity parameter that gives our best-fit is $\alpha \approx 1.5$ but if the spatial distribution of C I is affected by interactions with the planet or if the input rate is higher than the value used in our study (e.g. because of the inaccurate conversion of the CO flux to a CO mass 
and the possibility that other molecules contribute
to the production of C, O or even H), it could be as low as 0.1 (see subsection \ref{firstm}). This value is comparable to those found in highly ionised accretion discs \citep[e.g. dwarf novae in outburst;][]{2007MNRAS.376.1740K} in which the magnetorotational instability is thought to be operational. 
The condition for the ideal MRI to be operational is usually taken to be that the magnetic Reynolds number $Re_M$, given by
\begin{equation}
\label{Rem}
Re_M  = \frac{c_s H}{\eta},
\end{equation}

\noindent where $\eta$ is the magnetic diffusivity, exceeds some critical value $Re_M({\rm crit})$ (\citet{2000ApJ...530..464F} or \citet{2012MNRAS.420.3139M} suggest that 
$Re_M(crit) \approx 10^4$). In Figure \ref{Knudsenplot} (right), we plot $Re_M$ against radius for our best-fit accretion disc,  which shows that the magnetic Reynolds number exceeds this critical value by a large amount throughout the disc. 
We conclude that the value of $\alpha$ that we obtain is in line with the idea that the viscosity is provided by MHD turbulence (MRI).

\section{Discussion}

This paper presents a new model for gas in debris discs. Previous models had considered a static gas disc, whereas the model presented here couples its dynamical and thermal evolution, as well as taking into account its optical thickness to ionising
radiation. This allows for a better understanding of gas in debris discs. One of the main outcomes is that the various observations of gas around $\beta$ Pic can be explained within the framework of a viscous
evolution model. The model assumes that CO is produced from solid bodies and quickly photodissociates into C and O that evolve viscously. 
To explain the $\beta$ Pic carbon and CO observations in a self-consistent way, our gas model requires a CO mass input rate equal to
$\dot{M} \sim 1.4 \times 10^{18}$ kg, an $\alpha \sim 1.5$ for the viscous evolution and an impinging UV flux $F_i/F_0 \sim$ 60 times greater than the interstellar radiation field.

The required value of $\dot{M}$ is consistent with the CO input rate derived from observations \citep{2014Sci...343.1490D}. 

The $\alpha$ value needed to fit the carbon observations is rather high. We suggest that MRI is at work within this highly ionised medium, which seems to agree with the high magnetic Reynolds number that is found by our model. 
Hence, this work may provide the first indirect measurement of $\alpha$ in a debris disc. We note that the $\alpha$ value derived here is degenerate with $\dot{M}$ (as seen in subsection \ref{xhi2}) so that if $\dot{M}$ is greater by a factor 3, $\alpha$ would go down to $\sim$ 0.5. This value depends largely on the C I non-detection in our model and can be refined when the C I flux is known. Due to the uncertainties on C I and the uncertainties in our model, for now we put a lower limit on $\alpha$, which must be greater than 0.1.
The high $\alpha$ value found in our study is consistent with that inferred for dwarf novae and X-ray outbursts, where discs are fully ionised, which suggest $\alpha$ values ranging from 
0.1 to 0.4 \citep{2007MNRAS.376.1740K} and also with the value of $\alpha=1\pm0.2$ measured for the fully ionised disc in a Be star \citep{2012ApJ...744L..15C}. Note that the MRI conditions required to be active can involve some complicated non-linear effects (and some non ideal MRI effects), a discussion of which exceeds the scope of this paper \citep[see][for more details]{kral16c}.

We find that the overall UV flux (star+external radiation) impinging on the gas disc required to explain the observations is of the order of 60 times greater than the UV flux expected from the local interstellar radiation field. This is surprising, but $\beta$ Pic is still young and seems very active. 
Usually, mid-A to mid-B type stars are known to be X-ray dark but $\beta$ Pic shows some weak X-ray emission that could be coming from the thermal emission of a cool corona \citep{2012ApJ...750...78G}. Also, observations with
FUSE in $\beta$ Pic show the presence of highly ionised elements such as C III (977 \AA) and O VI (1032 \AA) pointing to the possible presence of an extended chromosphere or accretion around $\beta$ Pic \citep{2001ApJ...557L..67D}. We recall that in protoplanetary discs, the main ionisation source is the X-rays produced in T Tauri stars due to corona activity \citep{2011ApJ...739...50B}. These X-rays are not present around main sequence stars but there could be some
remnants in the youngest systems \citep{2015ApJ...801...31R}. Indeed, the flux of a typical classical T Tauri star FUV radiation field at 1 AU from the central star is $10^7$ times the interstellar radiation field \citep{2014ApJ...784..127F}. 

Also, the interstellar radiation field impinging on the $\beta$ Pic system is not well known. 
The IRF consists of four components that are described in Table A1 of \citet{1983A&A...128..212M} based on the well-observed spectrum of the IRF in the solar neighbourhood. 
The close-by environment of $\beta$ Pic could be modelled to take account of the presence of nearby O and B stars, but this represents an arduous
task \citep[e.g.][]{1973ApJ...181..363W} and was not attempted in this study. The IRF can also differ from that assumed in our model as the extinction curves of grains in the galaxy vary from one place
to another \citep{1994ASPC...58..319V}. Another uncertainty comes from the stellar spectrum.   
Overall, $\beta$ Pic shows a very active environment, which translates as increasing the ionising flux from the star in our model or the value of $F_i/F_0$. Having a more active star than usual main sequence A stars could explain why we need such a high radiation flux to fit the observations.

Previous work explaining the C II emission line observed towards $\beta$ Pic discarded the possibility of a viscous evolution because the emission line shows a very steep velocity gradient implying an absence of material in the inner region \citep{2014A&A...563A..66C}. 
We showed that this is not the case as viscous evolution does not create a continuous C II profile but rather one with a break resulting in a lower amount of C II in the inner regions. Indeed, the continuum optical thickness to ionising radiation in the inner regions blocks photoionisation, which in turn lowers the ionisation fraction and prevents 
C II from accumulating too much in the inner regions (its radial profile scales as $R^{-1.15}$). Thus, in our model, viscous evolution naturally creates the steep velocity gradient observed for the C II emission line.

\section{Summary-Conclusions}

We have developed a gas evolution model that couples the dynamics of gas particles to their thermal state through a viscous evolution that is modelled by an $\alpha$ prescription. Using a thermodynamical self-consistent model, we are able to
follow the evolution of the gas from its production site to its steady state location. This gives interesting insights on $\beta$ Pic and on the magnetorotational instability in general that can be summed up as follows:

\begin{itemize}
\item The model developed in this paper indicates that the dynamical evolution of gas in $\beta$ Pic is well represented through an $\alpha$ model where CO photodissociation produces C and O which diffuse viscously.
 \item The $\beta$ Pic gas disc is well explained by a viscous evolution with an $\alpha$ value greater than 0.1, a mass input rate of 0.1 M$_\oplus$/Myr (as found by the ALMA CO observation) and a high impinging UV flux (see Table~\ref{tab2}).
 \item $\beta$ Pic carbon observations are reproduced with our model assuming that it comes from CO. This model also explains the APEX non-detection of C I presented here and is in agreement with the 
O I detection by Herschel and the electron density derived from the CO J=2-1 and J=3-2 ratio.
 \item We make predictions for the hydrogen content in the $\beta$ Pictoris gas disc. We suggest that the H$_2$O/CO ratio in the colliding planetesimals is $\sim$ 1.5 (giving a predicted total H$_2$O mass present in the system of $\sim$ $2 \times 10^{-9}$ M$_\oplus$) and that the total H I column density along the line-of-sight is $\sim 3\times 10^{18}$ cm$^{-2}$, with a total H I mass of $3.1 \times 10^{-3}$ M$_\oplus$ (see Table~\ref{tab3} and \ref{tab4}).
 \item The unexpected observed X-ray flux in $\beta$ Pic may also be explained with our model, which provides the right amount of accretion to explain the X-ray observation.
  \item The ionisation fraction of carbon is high in $\beta$ Pic ($>0.3$). The ionisation fraction drops closer to the host star as the carbon density increases, which blocks more FUV flux from reaching the midplane.
 \item Due to the high ionisation fraction and high Reynolds magnetic number throughout the disc, we suggest that the magnetorotational instability is likely to be the physical mechanism that sets the viscosity in the disc \citep{kral16c}.
  \item We show that gas drag may be strong enough to affect the small unbound dust grains in $\beta$ Pic. This may help to explain the presence of small grains in the inner regions of $\beta$ Pic observed with GPI or the mid-IR emission within 20AU.
 \item Photoelectric heating is never dominant throughout the whole gas disc compared to photoionisation of carbon. The conditions for when dust heating can become important compared to carbon photoionisation heating can be estimated from Eq.~\ref{dustcond}.
\end{itemize}

Our model could be applied to other debris discs to make predictions for C I, C II, O I, H I line fluxes and so assess their observability. 


\section*{Acknowledgments}
We thank the referee for his/her helpful review. QK, MW and LM acknowledge support from the European Union through ERC grant number 279973. A.J. acknowledges the support of the DISCSIM project, grant agreement 341137, funded by the European Research Council under ERC-2013-ADG. 
QK wishes to thank Gianni Cataldi for providing
the C II emission line spectrum for $\beta$ Pictoris. This publication is based on data acquired with the Atacama Pathfinder Experiment (APEX), under program ID 096.F-9328(A). APEX is a collaboration between the Max-Planck-Institut fur Radioastronomie, 
the European Southern Observatory, and the Onsala Space Observatory. {\it Herschel} is an ESA space observatory with science instruments provided by European-led Principal Investigator consortia and with important participation from NASA.

\label{lastpage}

\end{document}